\documentclass{emulateapj} 
 
\usepackage[dvipdf]{epsfig} 
\usepackage[dvips]{rotating}
\usepackage{subfigure}
\usepackage{mathrsfs}
\usepackage{amsmath}

\newcommand{\IRAS}{{\it IRAS}}

\newcommand{\PLANCK}{{\it Planck}}
\newcommand{\AKARI}{{\it AKARI}}
\newcommand{\COBE}{{\it COBE}}
\newcommand{\WISE}{{\it WISE}}

\bibpunct{(}{)}{;}{a}{}{,} 
 
\shorttitle{\PLANCK~Two-component Dust Model}
 
\shortauthors{Meisner \& Finkbeiner} 

\begin{document}

\title{Modeling thermal dust emission with two components: application to the 
{\it PLANCK} HFI maps}
\author{Aaron M. Meisner\altaffilmark{1,2}}
\author{Douglas P. Finkbeiner\altaffilmark{1,2}}
\altaffiltext{1}{Department of Physics, Harvard University, 17 Oxford Street, 
Cambridge, MA 02138, USA; ameisner@fas.harvard.edu}
\altaffiltext{2}{Harvard-Smithsonian Center for Astrophysics, 60 Garden St, 
Cambridge, MA 02138, USA; dfinkbeiner@cfa.harvard.edu}

\begin{abstract}

We apply the \cite{FDS99} two-component thermal dust emission model to the
\PLANCK~HFI maps. This parametrization of the far-infrared dust spectrum as the
sum of two modified blackbodies serves as an important alternative to the 
commonly adopted single modified blackbody (MBB) dust emission model. Analyzing
the joint \PLANCK/DIRBE dust spectrum, we show that two-component models 
provide a better fit to the 100-3000 GHz emission than do single-MBB models, 
though by a lesser margin than found by \cite{FDS99} based on FIRAS and DIRBE. 
We also derive full-sky $6.1'$ resolution maps of dust optical depth and 
temperature by fitting the two-component model to \PLANCK~217-857 GHz along 
with  DIRBE/\IRAS~100$\mu$m data. Because our two-component model
matches the dust spectrum near its peak, accounts for the spectrum's flattening
at millimeter wavelengths, and specifies dust temperature at 6.1$'$ FWHM, our 
model provides reliable, high-resolution thermal dust emission foreground 
predictions from 100 to 3000 GHz. We find that, in diffuse sky regions, our 
two-component 100-217 GHz predictions are on average accurate to within 2.2\%, 
while extrapolating the \cite{planckdust} single-MBB model systematically 
underpredicts emission by 18.8\% at 100 GHz, 12.6\% at 143 GHz and 7.9\% at 217
GHz. We calibrate our two-component optical depth to reddening, and compare 
with reddening estimates based on stellar spectra. We find the dominant 
systematic problems in our temperature/reddening maps to be zodiacal light on 
large angular scales and the cosmic infrared background anisotropy on small 
angular scales.

\end{abstract}


\section{Introduction}
The presence of Galactic interstellar dust affects
astronomical observations over a wide range of wavelengths. In the mid-infrared
and far-infrared, Galactic dust emission contributes significantly
to the total observed sky intensity. At optical and ultraviolet (UV) 
wavelengths, dust grains absorb and scatter starlight. Observations of 
interstellar dust emission/absorption can improve our understanding of the 
physical conditions and composition of the interstellar medium (ISM), an 
environment which plays a crucial role in Galactic evolution and star 
formation. Equally, or perhaps even more important to the practice of 
astronomy, however, is accurately accounting for dust as a foreground which 
reddens optical/UV observations of stars/galaxies and superimposes Galactic 
emission on low-frequency observations of the cosmic microwave background 
(CMB).

Over the past decades, satellite observations have dramatically enhanced our
knowledge about infrared emission from the ISM. The \textit{Infrared Astronomy 
Satellite} (\IRAS), with its $\sim$4$'$ resolution, revolutionized the study of
Galactic dust emission, first revealing the high-latitude ``infrared cirrus'' 
using 60$\mu$m and 100$\mu$m observations \citep{low84, wheelock94} and 
highlighting the importance of detailed dust mapping in the 
far-infrared/submillimeter as a key foreground for cosmology. Later, the 
Diffuse Infrared Background Experiment (DIRBE) aboard the \COBE~satellite 
provided complementary full-sky measurements at ten infrared wavelengths from 
1.25$\mu$m to 240$\mu$m, boasting a reliable zero point despite inferior 
$\sim$0.7$^{\circ}$ angular resolution \citep{boggess92}. \COBE/FIRAS 
\citep{firas} also provided full-sky infrared dust spectra at $7^{\circ}$ 
resolution in 213 narrow frequency bins between 30 GHz and 2850 GHz.

\citet[hereafter FDS99]{FDS99} used these FIRAS data to derive a globally 
best-fit model of dust emission applicable over a very broad range of 
frequencies. FDS99 showed that no model consisting of a single modified 
blackbody (MBB) could accurately match the FIRAS/DIRBE spectrum at both the 
Wien and Rayleigh-Jeans extremes. To fit the thermal dust spectrum between 100 
and 3000 GHz, FDS99 therefore proposed an emission model consisting of two 
MBBs, each with a different temperature and emissivity power law index. 
Physically, these two components might represent distinct dust grain species 
within the ISM, or they might simply provide a convenient fitting function. By 
combining this best-fit two-component model with a custom reprocessing of 
DIRBE and \IRAS~100$\mu$m data, FDS99 provided widely used foreground 
predictions with $6.1'$ FHWM, limited largely by their $1.3^{\circ}$ resolution
DIRBE-based temperature correction.

The \PLANCK~2013 data release \citep{planck2013} represents an important 
opportunity to revisit foreground predictions in light of \PLANCK's superb, 
relatively artifact-free broadband data covering the entire sky and a wide 
range of frequencies. Towards this end, \cite{planckdust} has conducted a study
modeling \PLANCK~353 GHz, 545 GHz, 857 GHz and DIRBE/\IRAS~100$\mu$m emission 
with a single-MBB spectrum. More recently, \cite{aniano} has applied the 
\cite{dl07} dust grain model to \PLANCK, \IRAS, and \WISE~emission between 353 
GHz and $12\mu$m. Here we investigate the FDS99 two-component dust emission 
model as an alernative parametrization for the 100-3000 GHz dust spectral
energy distribution (SED) composed  of \PLANCK~High Frequency Instrument (HFI),
DIRBE and \IRAS~data. In doing so, we obtain \PLANCK-based maps of dust 
temperature and optical depth, both at $6.1'$ resolution. Because we employ a 
model that has been validated with FIRAS down to millimeter wavelengths and 
optimized for \PLANCK, our derived parameters are useful in constructing 
high-resolution predictions of dust emission over a very broad range of 
wavelengths. This includes low frequencies (100-350 GHz), which 
\cite{planckdust} caution their model may not adequately fit, and also 
wavelengths near the peak of the dust SED, relevant to e.g. 
\AKARI~140-160$\mu$m \citep{akari}. We also anticipate our derived optical 
depth map will serve as a valuable cross-check for extinction estimates based 
directly upon optical observations of stars \citep[e.g.][]{schlafly14} and as a
baseline for next-generation dust extinction maps incorporating 
high-resolution, full-sky infrared data sets such as 
\WISE~\citep{wright10, meisner14} and \AKARI.

In $\S$\ref{sec:data} we introduce the data used throughout this study. In 
$\S$\ref{sec:prepro} we describe our preprocessing of the \PLANCK~maps to 
isolate thermal emission from Galactic dust. In $\S$\ref{sec:modeling} we 
explain the two-component emission model we apply to the \PLANCK-based dust 
SED. In $\S$\ref{sec:bpcorr}, we discuss the details of predicting 
\PLANCK~observations based on this dust model. In $\S$\ref{sec:global} we 
derive constraints on our model's global parameters in light of the \PLANCK~HFI
maps. In $\S$\ref{sec:fitting} we detail the Markov chain Monte Carlo (MCMC) 
method with which we have estimated the spatially varying parameters of our 
model. In $\S$\ref{sec:ebv} we calibrate our derived optical depth to reddening
at optical wavelengths. In $\S$\ref{sec:em_compare} we compare our 
two-component thermal dust emission predictions to those of \cite{planckdust}. 
In $\S$\ref{sec:release} we present the full-sky maps of dust temperature and 
optical depth we have obtained, and conclude in $\S$\ref{sec:conclusion}.

\section{Data}
\label{sec:data}
All \PLANCK~data products utilized throughout this work are drawn from the 
\PLANCK~2013 release \citep{planck2013}. Specifically, we have made use 
of all six of the zodiacal light corrected HFI intensity maps
\citep[\texttt{R1.10\_nominal\_ZodiCorrected},][]{planckzodi}. Our 
full-resolution (6.1$'$ FWHM) SED fits neglect the two lowest HFI frequencies, 
100 and 143 GHz, as these have FWHM of 9.66$'$ and 7.27$'$ respectively.

To incorporate measurements on the Wien side of the dust emission spectrum, 
we include 100$\mu$m data in our SED fits. In particular, we use the 
\citet[henceforth SFD]{SFD} reprocessing of DIRBE/\IRAS~100$\mu$m, which we 
will refer to as \verb|i100|, and at times by frequency as 3000 GHz. The 
\verb|i100| map has angular resolution of $6.1'$, and was constructed so as to 
contain only thermal emission from Galactic dust, with compact sources and 
zodiacal light removed, and its zero level tied to H\,\textsc{i}. We use the 
\verb|i100| map as is, without any custom modifications. 

In some of our FIR dust SED analyses which do not require high angular 
resolution, specifically those of $\S$\ref{sec:global}, $\S$\ref{sec:lores}, 
and $\S$\ref{sec:hier}, we also make use of the SFD reprocessings of DIRBE 
140$\mu$m (2141 GHz) and 240$\mu$m (1250 GHz).

\section{Preprocessing}
\label{sec:prepro}

The following subsections detail the processing steps we have applied to 
isolate Galactic dust emission in the \PLANCK~maps in preparation for SED
fitting.

\subsection{CMB Anisotropy Removal}
\label{sec:cmb}
We first addressed the CMB anisotropies before performing any of the 
interpolation/smoothing described in $\S$\ref{sec:ptsrc}/$\S$\ref{sec:smth}. 
The CMB anisotropies are effectively imperceptible upon visual inspection 
of \PLANCK~857 GHz, but can be perceived at a low level in \PLANCK~545 GHz, and
are prominent at 100-353 GHz relative to the Galactic emission
we wish to characterize, especially at high latitudes. To remove the CMB 
anisotropies, we have subtracted the Spectral Matching Independent Component 
Analysis \citep[SMICA,][]{smica} model from each of the \PLANCK~maps, 
applying appropriate unit conversions for the 545 and 857 GHz maps with native 
units of MJy/sr. Low-order corrections, particularly our removal of  Solar 
dipole residuals, are discussed in $\S$\ref{sec:zp}.

\subsection{Compact Sources}
\label{sec:ptsrc}
After subtracting the SMICA CMB model, we interpolate over compact sources, 
including both point sources and resolved galaxies. Removing compact sources at
this stage is important as it prevents contamination of compact-source-free 
pixels in our downstream analyses which require smoothing of the \PLANCK~maps. 
SFD carefully removed point sources and galaxies from the \verb|i100| map 
everywhere outside of $|b|$$<$$5^{\circ}$. We do not perform any further 
modifications of the \verb|i100| map to account for compact sources. To mask 
compact sources in the \PLANCK~217-857 GHz maps, we use the SFD compact source 
mask. At 100, 143 GHz we use the compact source masks provided by the 
\PLANCK~collaboration in the file \verb|HFI_Mask_PointSrc_2048_R1.10.fits|. 
Given our pixelization (see $\S$\ref{sec:pix}), 1.56\% of pixels are masked 
at 217-857 GHz (1.05\%, 1.02\% at 100, 143 GHz).

\subsection{Smoothing}
\label{sec:smth}
For our full-resolution model, we wish to simultaneously fit \verb|i100| along 
with the four highest-frequency \PLANCK~bands. To properly combine these maps, 
they must have the same point spread function (PSF). \verb|i100|, with its 
$6.1'$ symmetric Gaussian beam, has the lowest angular resolution of the 
relevant maps. To match PSFs, we have therefore smoothed each of the 
\PLANCK~maps under consideration to \verb|i100| resolution by considering each 
native \PLANCK~map to have a symmetric Gaussian beam and smoothing by the 
appropriate symmetric Gaussian such that the resulting map has a  $6.1'$ FWHM. 
The FWHM values we assign to the native \PLANCK~maps are taken from 
\cite{planckbeam}, and are listed in Table \ref{table:offs}.

\subsection{Molecular Emission}
\label{sec:mole}
Because the FIRAS spectra consist of many narrow frequency bins, FDS99 were
able to discard the relatively small number of frequency intervals contaminated
by strong molecular line emission. Unfortunately, while the \PLANCK~data 
considered in this study are of high angular resolution, the broad 
\PLANCK~bandpasses do not allow us to adopt the same approach as FDS99 in 
dealing with line emission. Instead, we must subtract estimates of the 
molecular line contamination from each \PLANCK~band in order to best isolate 
the thermal continuum we wish to characterize. The most prominent molecular 
line emission in the \PLANCK~bands of interest arises from the three lowest CO 
rotational transitions: J=1$\rightarrow$0 at 115 GHz, J=2$\rightarrow$1 at 230 
GHz and J=3$\rightarrow$2 at 345 GHz, respectively affecting the \PLANCK~100, 
217 and 353 GHz bands. The J=1$\rightarrow$0 line also imparts a signal upon 
\PLANCK~143 GHz, but at a negligible level, $\sim$1000$\times$ fainter relative
to the dust continuum than J=1$\rightarrow$0 at 100 GHz. More specifically,
the ratio of J=1$\rightarrow$0 intensity to thermal dust emission in 
\PLANCK~143 GHz is $\geq$0.001 for only $<$2\% of the sky.

To correct for molecular emission, we employ the \PLANCK~Type 3 CO data 
product, which boasts the highest S/N among the available full-sky CO maps 
based on the \PLANCK~HFI and Low Frequency Instrument (LFI) data 
\citep{planckco}. The native angular resolution of the Type 3 CO map is 5.5$'$.
We therefore begin by smoothing the raw Type 3 CO map to match the PSF of the 
smoothed \PLANCK~intensity maps we wish to correct for molecular emission.

We must apply the appropriate unit conversions to the Type 3 CO map before 
subtracting it from the \PLANCK~intensity maps, which have native units of 
$K_{CMB}$ at the frequencies of interest. The Type 3 CO map is provided in 
units of K$_{RJ}$ km/s of J=1$\rightarrow$0 emission. To convert this quantity 
to $K_{CMB}$, we assume that all of the CO emission arises from the $^{12}$CO 
isotope, and derive the \PLANCK-observed CO intensity in units of $K_{CMB}$ as
follows:

\begin{equation}
I_{CO, \nu_i, N, N-1} = I_{3}F_{12CO, \nu_i, N, N-1} R_{N, N-1}
\end{equation}

Where $I_{CO, \nu_i, N, N-1}$ is the intensity in $K_{CMB}$ in \PLANCK~band 
$\nu_i$ due to the CO transition from J=$N$ to J=$(N$$-$1). $I_3$ represents 
the appropriately smoothed Type 3 CO amplitude in  K$_{RJ}$ km/s of 
J=1$\rightarrow$0 emission. The $F_{12CO, \nu_i, N, N-1}$ are conversion 
factors between K$_{RJ}$ km/s and $K_{CMB}$ for particular band/transition 
pairs. The relevant values, calculated with the \textit{Unit Conversion and 
Colour Correction} software utilities (\verb|v1.2|), are:

\noindent
$F_{12CO, 100, 1, 0}$=1.478$\times$10$^{-5}$$K_{CMB}/(K_{RJ} \ km/s)$,

\noindent
$F_{12CO, 217, 2, 1}$=4.585$\times$$10^{-5}$$K_{CMB}/(K_{RJ} \ km/s)$, and 

\noindent
$F_{12CO, 353, 3, 2}$=1.751$\times$$10^{-4}$$K_{CMB}/(K_{RJ} \ km/s)$.

\noindent
$R_{N, N-1}$ represents the line ratio of the transition from J=$N$ to 
J=$(N$$-$1) relative to the J=1$\rightarrow$0. Thus, $R_{1,0}$=1, and we 
further adopt $R_{2,1}$=0.595 and $R_{3,2}$=0.297 based on \cite{planckco}. 
These line ratios are assumed to be constant over the entire sky. 

Formally, then, the CO contamination in band $\nu_i$ is given by:

\begin{equation} \label{equ:molcorr}
I_{CO, \nu_i} = \sum\limits_{N} I_{CO, \nu_i, N, N-1}
\end{equation}

It happens that, for each of the \PLANCK~bands in which CO emission is
non-negligible (100, 217 and 353 GHz), only a single $N$ contributes ($N$=1, 
$N$=2 and $N$=3, respectively). 

Unfortunately, the Type 3 CO map at $6.1'$ FWHM is rather noisy, and the vast
majority of the sky has completely negligible CO emission. Thus, in order to 
avoid adding unnecessary noise outside of molecular cloud complexes and at high
latitudes, we have zeroed out low-signal regions of the Type 3 CO map. We 
identify  low-signal regions as those with $\mathcal{I}_3$$<$1 K$_{RJ}$ km/s, 
where $\mathcal{I}_3$ is the Type 3 CO map smoothed to 0.25$^{\circ}$ FWHM. As 
a result of this cut, 90\% of the sky remains unaffected by our CO correction, 
particularly the vast majority of the high Galactic latitude sky.

\subsection{Zero Level}
\label{sec:zp}

Although we wish to isolate and model thermal emission from Galactic dust, the
\PLANCK~maps contain additional components on large angular scales. At each 
frequency, there can exist an overall, constant offset that must be subtracted 
to set the zero level of Galactic dust by removing the mean cosmic IR 
background \citep[CIB,][]{cibreview}, as well as any instrumental offset. 
Additionally, faint residuals of the Solar dipole remain at low frequencies. We
will address these issues by separately solving two sub-problems: first, we set
the absolute zero level of \PLANCK~857 GHz relative to external data, and 
second we fit the 100-545 GHz offsets and low order corrections by correlating 
these \PLANCK~bands against \PLANCK~857 GHz.

\subsubsection{Absolute Zero Level}
\label{sec:zp_abs}
In \cite{planckdust}, the absolute zero level of thermal dust emission was set 
by requiring that \PLANCK~infrared emission tends to zero when H\,\textsc{i} is
zero, assuming a linear correlation between these two measurements at low 
column density. However, this approach is less than completely satsifying in 
that there appear to be different slopes of \PLANCK~857 GHz versus 
H\,\textsc{i} for different ranges of H\,\textsc{i} intensity. In particular, 
\PLANCK~857 GHz appears to ``flatten out'' at very low H\,\textsc{i}, as shown
in Figure 5 of \cite{planckdust}. More quantitavely, we have found using the 
LAB H\,\textsc{i} data \citep{lab} for $-72$$<$$v_{LSR}$$<$$+25$ km/s that the 
best-fit slope for H\,\textsc{i}$<$70 K km/s is a factor of $\sim$1.9 lower 
than the best fit slope for 110 K km/s $<$H\,\textsc{i}$<$200 K km/s, and as a 
result the implied zero level offsets for \PLANCK~857 GHz differ by $\sim$0.37 
MJy/sr.

\begin{figure}
\begin{center}
\epsfig{file=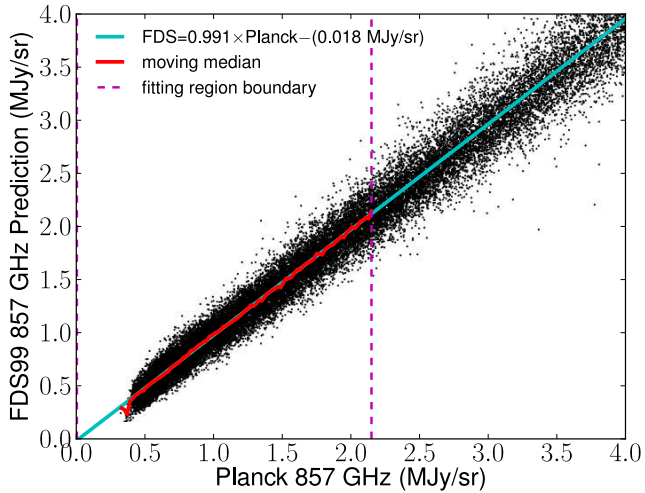, width=3.4in}
\caption{\label{fig:fdsref} Scatter plot of FDS99-predicted 857 GHz thermal
dust emission versus \PLANCK~857 GHz observations, illustrating our absolute 
zero level determination described in $\S$\ref{sec:zp_abs}.}
\end{center}
\end{figure}

\begin{figure}
\begin{center}
\epsfig{file=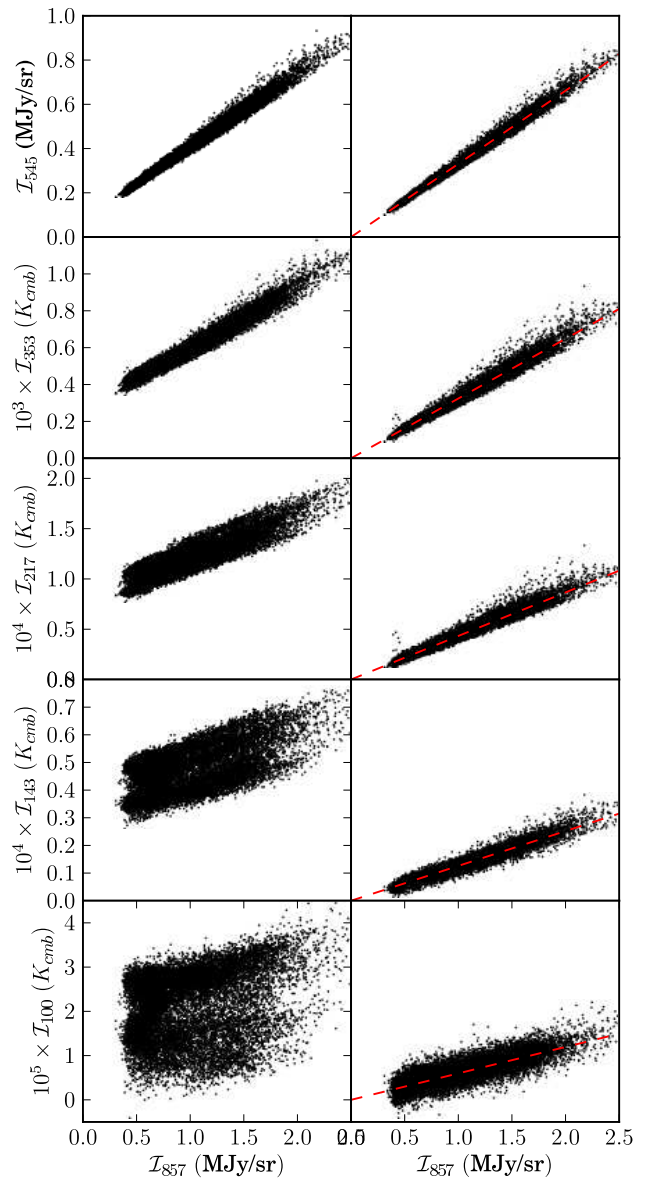, width=3.4in}
\caption{\label{fig:dip}  Scatter plots of \PLANCK~100, 143, 217, 353, and 545 
GHz versus \PLANCK~857 GHz. Left: before applying our best-fit zero level 
offsets and additional low-order corrections. Right, top four panels: 
\PLANCK~143-545 GHz after correcting for each band's best-fit offset and 
residual Solar dipole. Bottom right: \PLANCK~100 GHz after applying the 
spherical harmonic corrections of Equation \ref{equ:harm}. The dashed red line 
shows the best-fit linear relationship in all cases.}
\end{center}
\end{figure}

\begin{figure*}
\begin{center}
\epsfig{file=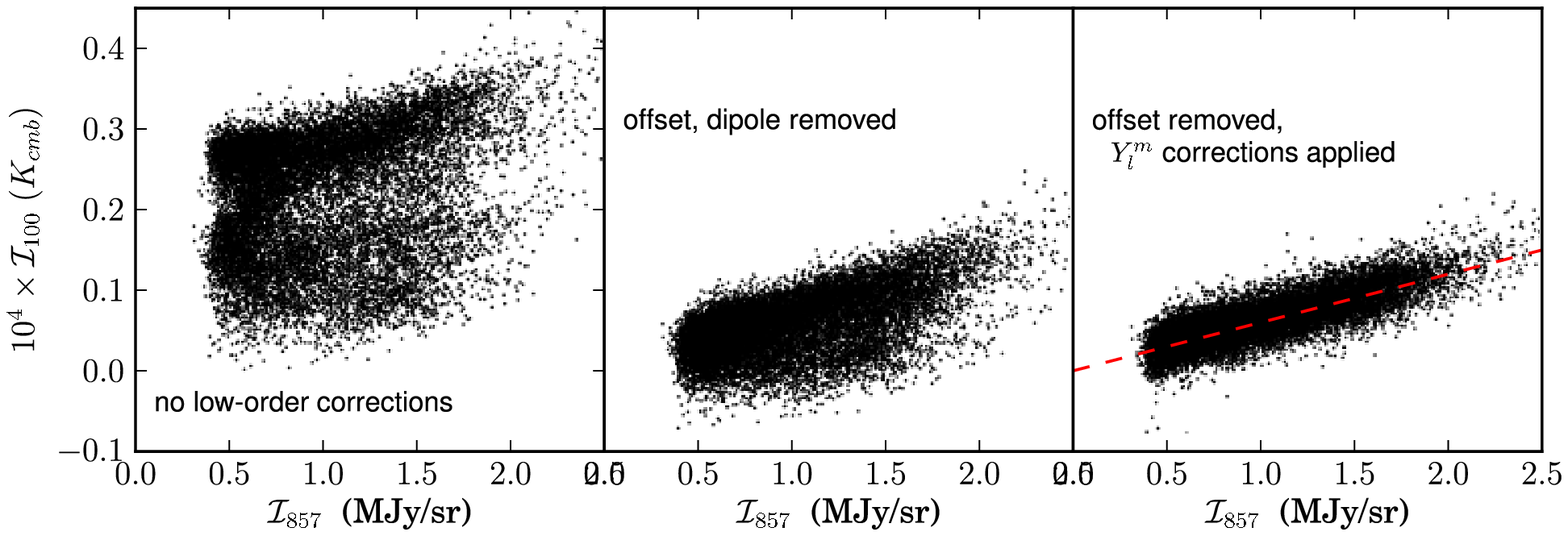, width=6.5in}
\caption{\label{fig:harm} Summary of low-order corrections at 100 GHz. Left: 
prior to our low-order corrections, a $\sim$17$\mu$K zero level offset is 
present and strong low-order problems reduce the linearity of the 100 GHz 
trend versus 857 GHz. Center: scatter plot of \PLANCK~100 GHz versus 857 GHz 
after applying the best-fit offset and residual Solar dipole corrections 
derived with Equation \ref{equ:dip} to \PLANCK~100 GHz. The correlation is 
strengthened, but remains far less tight than for 143-545 GHz (see right 
column of Figure \ref{fig:dip}, top four rows). Right: after applying the 
spherical harmonic corrections of Equation \ref{equ:harm} to \PLANCK~100 GHz, 
the correlation versus 857 GHz is far more tightly linear than following the 
dipole correction.}
\end{center}
\end{figure*}

Because of this ambiguity in the relationship between 857 GHz and H\,\textsc{i}
emission, we decided to instead constrain the \PLANCK~857 GHz zero level by 
comparison to the FDS99-predicted 857 GHz thermal dust emission. This renders 
our \PLANCK~857 GHz absolute zero level tied indirectly to H\,\textsc{i} 
through the FDS99 100$\mu$m and 240$\mu$m zero levels.

We perform a linear fit to the FDS99-predicted 857 GHz values as 
a function of \PLANCK~857 GHz. For this purpose, we employ a version of the 
\PLANCK~857 GHz map with zodiacal light and point sources removed and smoothed 
to 1$^{\circ}$ FWHM, which we will refer to as $\mathcal{I}_{857}$. We 
consider $\mathcal{I}_{857}$ to be the independent variable, as it has much 
higher S/N than the FDS99 prediction, henceforward referred to as 
$\mathcal{F}_{857}$. Note that $\mathcal{F}_{857}$ is not simply the FDS99
model evaluated at 857 GHz, but also incorporates the color correction factor 
of $\S$\ref{sec:bpcorr}, using the FDS99 temperature map to determine the dust 
spectrum shape. We rebin to $N_{side}$=64 and restrict to pixels with 
$\mathcal{I}_{857}$$<$2.15 MJy/sr. Since \PLANCK~857 GHz smoothed to degree 
resolution has very high S/N, we can safely perform such a cut on 
$\mathcal{I}_{857}$. Figure \ref{fig:fdsref} shows a scatter plot of 
$\mathcal{I}_{857}$ versus $\mathcal{F}_{857}$, with a moving median and linear
fit overplotted. The linear fit was performed with uniform weights and 
iterative outlier rejection. The best-fit linear model is given by 
$\mathcal{F}_{857}$=0.991$\mathcal{I}_{857}$$-$0.018 MJy/sr. It is encouraging 
that the slope is quite close to unity. It is also encouraging that our 
choice of \PLANCK~857 GHz threshold at 2.15 MJy/sr is unimportant; any 
threshold value between 1.3 MJy/sr (28$^{th}$ percentile in 
$\mathcal{I}_{857}$) and 3.9 MJy/sr (61$^{st}$ percentile in 
$\mathcal{I}_{857}$) yields a zero level offset within 0.01 MJy/sr of our 
adopted value.

The formal statistical error on the best-fit 857 GHz offset is quite small, 
$\sim$0.002 MJy/sr. The systematics likely to dominate the actual uncertainty 
on our FDS-based zero level are imperfections in the \PLANCK/\verb|i100| 
zodiacal light models and the FDS99 temperature map. To quantify these 
systematic uncertainties, we split the sky into four quadrants, with boundaries
at $b$=0$^{\circ}$ and $l$=0$^{\circ}$, $l$=180$^{\circ}$. We again restricted 
to $\mathcal{I}_{857}$$<$2.15 MJy/sr, and repeated the regression in each 
quadrant. The rms of the per-quadrant slopes was found to be 0.0188, while the 
rms of the per-quadrant offsets was 0.0586 MJy/sr. Our adopted $\sim$0.06 
MJy/sr zero level uncertainty is sufficiently large to be consistent with 
the possible error introduced by assuming no appreciable Solar dipole signal in
the \PLANCK~857 GHz map. If we allow for a dipole template in our FDS99 versus
\PLANCK~linear regression at 857 GHz, the best-fit dipole amplitude is only 
0.02 MJy/sr. 

\begin{deluxetable*}{ccccccccc}
\tabletypesize{\scriptsize}
\tablecolumns{8} 
\tablewidth{0pc} 
\tablecaption{\label{table:offs} Input Map Properties \& Pre-processing} 
\tablehead{
\colhead{$\nu$ (GHz)} &
\colhead{Instrument(s)} &
\colhead{Offset ($K_{CMB}$)} & 
\colhead{Dipole ($K_{CMB}$)} & 
\colhead{$s_{857,\nu}$$\times$$u_{\nu}$} &
\colhead{$\sigma_{s_{857,\nu}}$$\times$$u_{\nu}$} &
\colhead{$n_{\nu}$} ($K_{CMB}$) &
\colhead{$c_{\nu}$} &
\colhead{FWHM ($'$)}
}
\startdata
100  & \PLANCK~HFI & 1.69$\times$10$^{-5}$$\pm$3.61$\times$10$^{-7}$ & $-$1.08$\times$10$^{-5}$ & 1.46$\times$10$^{-3}$ & 2.92$\times$10$^{-5}$  & 7.77$\times$10$^{-5}$  & 0.0054 & 9.66 \\
143  & \PLANCK~HFI & 3.58$\times$10$^{-5}$$\pm$7.58$\times$10$^{-7}$ & $-$1.08$\times$10$^{-5}$ & 4.68$\times$10$^{-3}$ & 9.37$\times$10$^{-5}$  & 3.25$\times$10$^{-5}$  & 0.0054 & 7.27 \\
217  & \PLANCK~HFI & 7.79$\times$10$^{-5}$$\pm$2.60$\times$10$^{-6}$ & $-$1.40$\times$10$^{-5}$ & 2.09$\times$10$^{-2}$ & 4.19$\times$10$^{-4}$  & 4.51$\times$10$^{-5}$  & 0.0054 & 5.01 \\
353  & \PLANCK~HFI & 2.76$\times$10$^{-4}$$\pm$1.95$\times$10$^{-5}$ & $-$3.08$\times$10$^{-5}$ & 9.32$\times$10$^{-2}$ & 1.86$\times$10$^{-3}$  & 1.51$\times$10$^{-4}$  & 0.012  & 4.86 \\
     &             & Offset (MJy/sr)                                 & Dipole (MJy/sr)          & $s_{857,\nu}$         & $\sigma_{s_{857,\nu}}$ & $n_{\nu}$ (MJy/sr)     &        &      \\ \cline{3-7} \\ [-2ex]
545  & \PLANCK~HFI & 7.27$\times$10$^{-2}$$\pm$1.99$\times$10$^{-2}$ & 1.63$\times$10$^{-2}$    & 3.31$\times$10$^{-1}$ & 6.62$\times$10$^{-3}$  & 0.046                  & 0.10   & 4.84 \\
857  & \PLANCK~HFI & 1.82$\times$10$^{-2}$$\pm$6.02$\times$10$^{-2}$ & -                        & 1.0                   & 2.0$\times$10$^{-2}$   & 0.046                  & 0.10   & 4.63 \\
1250 & DIRBE       & 7.06$\times$10$^{-2}$$\pm$1.19$\times$10$^{-1}$ & -                        & 1.98                  & 3.97$\times$10$^{-2}$  & 0.42                   & 0.10   & 42   \\
2141 & DIRBE       & 1.04$\times$10$^{-1}$$\pm$1.54$\times$10$^{-1}$ & -                        & 2.56                  & 5.12$\times$10$^{-2}$  & 0.79                   & 0.10   & 42   \\
3000 & DIRBE/\IRAS & 0.0$\pm$4.3$\times$10$^{-2}$                    & -                        & 1.27                  & 2.53$\times$10$^{-2}$  & 0.06                   & 0.10  & \ 6.1 
\enddata
\tablecomments{Column 1: Approximate band center frequency of each input map. 
Note that 1250 GHz and 2141 GHz refer to the SFD98 reprocessings of DIRBE 
240$\mu$m and 140$\mu$m respectively. Column 2: Instrument(s) from which the 
input map at each frequency has been obtained. Column 3: Zero level offset 
subtracted from each raw input map. Column 4: Best-fit residual Solar dipole 
amplitude according to Equation \ref{equ:dip}. Column 5: Dimensionless
correlation slope of each map relative to \PLANCK~857 GHz. These are the 
correlation slopes used in the analysis of $\S$\ref{sec:global}, specifically 
Equation \ref{equ:chi2corr}. Column 6: Adopted uncertainty on the dimensionless
correlation slopes relative to \PLANCK~857 GHz, for use in Equation 
\ref{equ:chi2corr}. Column 7: $n_{\nu}$ represents the adopted per-pixel 
statistical noise level at full resolution, which contributes to the error 
budget of Equation \ref{equ:errorbudget}. Column 8: Multiplicative fractional 
uncertainty on each input map, for use in the error budget of Equation 
\ref{equ:errorbudget}. Column 9: Native angular resolution of each input map.}
\end{deluxetable*}

\subsubsection{Relative Zero Level}
\label{sec:relzero}

In the course of this study we use not only \PLANCK~857 GHz, but also all of
the remaining \PLANCK~HFI bands, as well as \verb|i100|. To derive the 
zero level offsets that must be applied to each of the five lowest-frequency 
\PLANCK~bands, we perform a regression versus the \PLANCK~857 GHz map corrected
for the best-fit absolute zero level offset from $\S$\ref{sec:zp_abs}. We 
assume no offset need be applied to \verb|i100|, which already has its zero 
level tied to H\,\textsc{i} by SFD.

The need for additional low-order corrections beyond simple scalar offsets 
became evident upon inspecting the HFI maps at 100-545 GHz. In particular, we 
noticed the presence of a low-level dipole pattern, with an orientation 
consistent with that of the Solar dipole. Our strategy will be to 
simultaneously fit both this residual dipole and the zero-level offset 
amplitude for each band. To most precisely recover these amplitudes, it is 
necessary to have the highest available S/N in the independent variable of our 
regression. For this reason we have used \PLANCK~857 GHz as a reference for the
100-545 GHz bands, as opposed to the FDS99 predictions or H\,\textsc{i} data. 
In doing so, we assume \PLANCK~857 GHz contains no appreciable Solar dipole 
residual.

We perform one regression per HFI band (other than 857 GHz) to simultaneously 
fit for the zero level offset, the slope relative to 857 GHz, and the residual 
dipole amplitude. For each 100-545 GHz HFI band, we restrict to regions of low 
column density (H\,\textsc{i} $<$ 200 K\,km\,s$^{-1}$ for 
$-72$$<$$v_{LSR}$$<$$+25$ km\,s$^{-1}$) and fit the following model:

\begin{equation} \label{equ:dip}
\mathcal{I}_{\nu_i, p} = m\mathcal{I}_{857, p} + b + d\mathcal{D}_{p}
\end{equation}

With $p$ denoting a single $N_{side}$=64 HEALPix pixel \citep{healpix} in the 
maps $\mathcal{I}_{857}$, $\mathcal{I}_{\nu_i}$, and $\mathcal{D}$. Here 
$\mathcal{I}_{857}$ is the \PLANCK~857 GHz map with zodiacal emission, compact 
sources, and the constant offset of $\S$\ref{sec:zp_abs} removed, smoothed to 
$1^{\circ}$ resolution. $\mathcal{I}_{\nu_i}$ is the corresponding $1^{\circ}$ 
resolution \PLANCK~HFI map with zodiacal emission, CMB anisotropies, and
compact sources removed. In the context of Equation \ref{equ:dip}, $\nu_i \in$ 
\{100, 143, 217, 353, 545\} GHz. Note that $\mathcal{I}_{\nu_i}$ is always in
the native units of the relevant \PLANCK~band. $\mathcal{D}$ is a scaling of 
the Solar dipole pattern oriented toward 
$(l, b) = (263.99^{\circ}, 48.26^{\circ})$, with unit amplitude. Because 
$\sim$18,000 pixels satisfy the low H\,\textsc{i} cut, we have an 
overconstrained linear model with three parameters: $m$, $d$, and $b$. $m$ 
represents the best-fit slope of \PLANCK~band $\nu_i$ versus \PLANCK~857 GHz 
assuming they are linearly related. $d$ is the residual Solar dipole amplitude,
and its best-fit value represents the scaling of the Solar dipole that makes 
the \PLANCK~band $\nu_i$ versus 857 GHz correlation most tightly linear. $b$ 
represents the constant offset that must be subtracted from the band $\nu_i$ 
map to make its zero level consistent with that of the 857 GHz map. For each 
band $\nu_i$, we obtain estimates of $m$, $d$, and $b$ by performing a linear 
least squares fit with uniform weights and iterative outlier rejection. Figure 
\ref{fig:dip} shows scatter plots of the band $\nu_i$ versus 857 GHz 
correlation before (left) and after (right) correcting for the best-fit offset 
and residual dipole, for each $\nu_i \in$ \{143, 217, 353, 545\} GHz. Not only 
are the tightened correlations striking in these scatter plots, but the 
residual dipole subtractions appear very successful in the two-dimensional band
$\nu_i$ maps themselves. Before performing thermal dust fits, we therefore 
subtract the best-fit $b$ and $d\mathcal{D}$ from each 143-545 GHz map. The 
best-fit offsets and residual dipole amplitudes are listed in Table 
\ref{table:offs}, along with other important per-band parameters, such as the 
fractional multiplicative calibration uncertainty $c_{\nu}$.

We found that a dipole correction alone could not sufficiently rectify the 
\PLANCK~100 GHz map (see Figure \ref{fig:harm}). Therefore, for 100 GHz, we 
performed a modified version of the Equation \ref{equ:dip} fit, using the 
following model:

\begin{equation} \label{equ:harm}
\mathcal{I}_{100, p} = m\mathcal{I}_{857, p} + b + \sum_{l=1}^{4} \sum_{m=-l}^{l} a_{lm}Y_{l}^{m}(\theta_p, \phi_p)
\end{equation}

Where $Y_{l}^{m}$ are the real spherical harmonics, and the $a_{lm}$ are their 
corresponding real coefficients. The angle $\phi_p$ is taken simply to be 
$l_{gal, p}$ and $\theta_p$=$(90^{\circ}-b_{gal, p})$. Thus, we have replaced 
the Solar dipole term with a sum of 24 spherical harmonic templates, which, 
when multiplied by the best-fit $a_{lm}$ coefficients and subtracted from 
\PLANCK~100 GHz make the relation between 100 GHz and 857 GHz most tightly 
linear. Figure \ref{fig:harm} illustrates the improved correlation of 100 GHz 
vs. 857 GHz when including the spherical harmonic corrections relative to the 
dipole-only correction. The spherical harmonic decomposition of Equation 
\ref{equ:harm} did not improve the correlations at higher frequencies 
enough to warrant replacing the dipole-only correction in those cases.

\section{Dust Emission Model}
\label{sec:modeling}

At sufficiently high frequencies, Galactic thermal dust emission can be 
adequately modeled as a single MBB with power-law emissivity 
\citep[e.g. SFD;][]{planckdust}. However, it has long been recognized, 
particularly in view of the FIRAS spectra, that the dust SED flattens toward 
the millimeter in a manner which is not consistent with a simple extrapolation 
of single-MBB models to low frequencies. In the diffuse ISM, \cite{reach95} 
found an improved fit to the FIRAS data using an empirically motivated 
superposition of two $\beta$=2 MBBs, one representing a `hot' grain population 
($T$$\approx$16$-$21 K), the other a `cold' grain population 
($T$$\approx$4$-$7 K). FDS99 built a more physically motivated two-MBB model, 
in which different grain emission/absorption properties account for the 
differing temperatures of each population, and these temperatures are coupled 
by assuming thermal equilibrium with the same interstellar radiation field 
(ISRF).

The primary FDS99 analysis considered the intrinsic grain properties of each
species, for example the emissivity power law indices, to be constant over the 
sky, and performed a correlation slope analysis to constrain these parameters 
with FIRAS and DIRBE observations. FDS99 also constructed a DIRBE 
240$\mu$m/100$\mu$m ratio to account for temperature variation at 
$\sim$1.3$^{\circ}$ resolution. In this work we seek to apply the FDS99 
emission model to the \PLANCK~data set, which offers a dramatic enhancement in
angular resolution relative to the FIRAS spectra. The \PLANCK~data thereby 
allow us to derive an improved temperature correction at near-\IRAS~resolution
($\S$\ref{sec:mcmc}), re-evaluate the best-fit global dust properties 
($\S$\ref{sec:global}, $\S$\ref{sec:hier}), and fit additional 
two-component model parameters as a function of position on the sky 
($\S$\ref{sec:lores}).

The shape of the two-component model spectrum we will consider is given by:

\begin{equation}
M_{\nu} \propto \Big[f_{1}q_{1}\Big(\frac{\nu}{\nu_{0}}\Big)^{\beta_1}B_{\nu}(T_1) + f_{2}q_{2}\Big(\frac{\nu}{\nu_0}\Big)^{\beta_2}B_{\nu}(T_2)\Big]
\end{equation}

Where $B_{\nu}$ is the Planck function, $T_1$ is the `cold' dust temperature, 
$T_2$ is the `hot' dust temperature, and $\beta_1$ and $\beta_2$ are the
emissivity power-law indices of the cold and hot dust components respectively. 
$q_1$ represents the ratio of FIR emission cross section to optical absorption 
cross section for species 1, and similarly $q_2$ for species 2. $f_1$ and 
$f_2$ dictate the relative contributions of the two MBB components to the
combined SED. Thus, $f_1$ and $f_2$ can be thought of as encoding the mass 
fraction of each species, although technically $f_1$ ($f_2$) is the optical 
absorption cross-section weighted mass fraction for species 1 (2). Following 
the convention of FDS99, we choose $\nu_0$=3000 GHz and take $f_2$=(1$-$$f_1$).

Mathematically, this two-MBB model requires specification of seven 
parameters for every line of sight: $T_1$, $T_2$, $\beta_1$, $\beta_2$, $f_1$, 
$q_1$/$q_2$ and the normalization of $M_{\nu}$. However, under the 
assumption that the temperature of each species is determined by maintaining 
thermal equilibrium with the same ISRF, $T_1$=$T_1$($T_2$, $\beta_1$, 
$\beta_2$, $q_1/q_2$) is fully determined by these other parameters. $T_1$ is 
always related to $T_2$ via a simple power law, although the prefactor and 
exponent depend on the parameters $q_1/q_2$, $\beta_1$ and $\beta_2$ (see FDS99
Equation 14).

These considerations still leave us with six potentially free parameters per 
line of sight. Unfortunately, fitting this many parameters per spatial pixel is
not feasible for our full-resolution $6.1$$'$ fits, as these are constrained by
only five broadband intensity measurements. Hence, as in FDS99, we deem certain
 parameters to be ``global'', i.e. spatially constant over the entire sky. In 
our full-resolution five-band fits, we designate $\beta_1$, $\beta_2$, $f_1$ 
and $q_1/q_2$ to be spatially constant. This same approach was employed by 
FDS99, and the globally best-fit values obtained by FDS99 for these parameters 
are listed in the first row of Table \ref{tab:global}. With these global 
parameters, FDS99 found $T_2$$\approx$$16.2$K, $T_1$$\approx$$9.4$K to be 
typical at high-latitude. In $\S$\ref{sec:global}, we discuss the best-fit 
global parameters favored by the \PLANCK~HFI data; these are listed in the 
second row of Table \ref{tab:global}.

\begin{deluxetable*}{llrrrrrrrrrr} 
\tabletypesize{\scriptsize}
\tablecolumns{11} 
\tablewidth{0pc} 
\tablecaption{\label{tab:global} Global Model Parameters} 
\tablehead{
\colhead{Number} &
\colhead{Model} &
\colhead{$f_1$} &
\colhead{$q_1/q_2$} & 
\colhead{$\beta_1$} & 
\colhead{$\beta_2$} &
\colhead{$T_2$} &
\colhead{$T_1$} &
\colhead{$n$} &
\colhead{D.O.F.} &
\colhead{$\chi^2$} &
\colhead{$\chi^2_{\nu}$}
}
\startdata
 1 & FDS99 best-fit  & 0.0363 & 13.0  & 1.67 & 2.70 & 15.72 &  9.15 & 1.018 & 7 & 23.9 & 3.41 \\
 2 & FDS99 general   & 0.0485 & 8.22  & 1.63 & 2.82 & 15.70 &  9.75 & 0.980 & 3 & 3.99 & 1.33 \\
 3 & single MBB      &  0.0   &  ...  &  ... & 1.59 & 19.63 &   ... & 0.999 & 6 & 33.9 & 5.65 \\ [-2ex]
\enddata
\end{deluxetable*}

Fixing the aforementioned four global parameters, our full-resolution, 
five-band fits have two remaining free parameters per line of sight: the hot 
dust temperature $T_2$ determines the SED shape and the normalization of 
$M_{\nu}$ determines the SED amplitude. In the lower-resolution fits of 
$\S$\ref{sec:lores} which include all HFI bands, we will allow $f_1$ to be a 
third free parameter, still holding $\beta_1$, $\beta_2$, and $q_1/q_2$ fixed.

To calculate the optical depth in the context of this model, we assume
optically thin conditions, meaning that $\tau_{\nu}$ = $M_{\nu}/S_{\nu}$, where
$M_{\nu}$ is the appropriately scaled two-component model intensity and the 
source function is given by:

\begin{equation}
\label{eqn:source}
S_{\nu} = \frac{f_1q_1(\nu/\nu_0)^{\beta_1}B_{\nu}(T_1) + f_2q_2(\nu/\nu_0)^{\beta_2}B_{\nu}(T_2)}{f_1q_1(\nu/\nu_0)^{\beta_1}+f_2q_2(\nu/\nu_0)^{\beta_2}}
\end{equation}

\section{Predicting the Observed SED}
\label{sec:bpcorr}

The thermal dust emission model of $\S$\ref{sec:modeling} predicts the 
flux density per solid angle $M_{\nu}$ in e.g. MJy/sr for any single frequency 
$\nu$. In practice, however, we wish to constrain our model using measurements 
in the broad \PLANCK/DIRBE bandpasses, each with $\Delta\nu/\nu\sim0.3$. 
Both the \PLANCK~and DIRBE data products quote flux density per solid 
angle in MJy/sr under the `IRAS convention'. More precisely, each value 
reported in the \PLANCK~maps gives the amplitude of a power-law spectrum 
with $\alpha$=$-1$, evaluated at the nominal band center frequency, such that 
this spectrum integrated against the transmission reproduces the 
bolometer-measured power. Because our model spectra do not conform to the 
$\alpha$=$-1$ convention, we have computed color correction factors to account 
for the MBB($T$, $\beta$) spectral shape and the transmission as a function of 
frequency:

\begin{equation} \label{equ:bpcorr}
b_{\nu_i}(T, \beta) = \frac{\int \nu^{\beta}B_{\nu}(T)\mathcal{T}_{\nu_i}(\nu) d\nu \bigg[\int (\nu_{i,c}/\nu)\mathcal{T}_{\nu_i}(\nu) d\nu\bigg]^{-1}}{\nu_{i,c}^{\beta}B_{\nu_{i,c}}(T)}
\end{equation}

Here $\nu_{i,c}$ is the nominal band center frequency of band $\nu_i$,  with 
$\nu_{i,c} \in$ \{100, 143, 217, 353, 545, 857, 1249.1352, 2141.3747, 
2997.92458\} GHz. $\mathcal{T}_{\nu_i}(\nu)$ represents the relative 
transmission as a function of frequency for band $\nu_i$. For the HFI maps, 
$\mathcal{T}_{\nu_i}(\nu)$ is given by the \PLANCK~transmission curves provided
in the file \verb|HFI_RIMO_R1.10.fits| \citep{planckresponse}. For \verb|i100| 
and DIRBE 140$\mu$m, 240$\mu$m, we have adopted the corresponding DIRBE 
transmission curves. 

The two-component model prediction in band $\nu_i$ under the IRAS convention, 
termed $\tilde{I}_{\nu_i}$, is then constructed as a linear combination of 
color-corrected MBB terms:

\begin{equation} \label{equ:iras}
\tilde{I}_{\nu_i} \propto \sum_{k=1}^{2} b_{\nu_i}(T_k, \beta_k) f_k q_k (\nu_{i,c}/\nu_0)^{\beta_k} B_{\nu_{i,c}}(T_k)
\end{equation}

The color correction of Equation \ref{equ:bpcorr} therefore allows us to 
predict $\tilde{I}_{\nu_i}$ by computing monochromatic flux densities at the 
central frequency $\nu_{i,c}$ and then multiplying by factors 
$b_{\nu_i}(T, \beta)$. In practice, we interpolated the color corrections off 
of a set of precomputed, one-dimensional lookup tables each listing 
$b_{\nu_i}(T, \beta)$ for a single $\beta$ value as a function of $T$. We thus
avoided the need to interpolate in both $\beta$ and $T$ by computing only a 
small set of one dimensional correction factors for the particular set of 
$\beta$ values of interest (e.g. $\beta$=1.67, 2.7, 1.63, 2.82 ..., see Table 
\ref{tab:global}). This color correction approach makes the MCMC sampling 
described in $\S$\ref{sec:mcmc} much more computationally efficient by 
circumventing the need to perform the integral in the numerator of Equation 
\ref{equ:bpcorr} on-the-fly for each proposed dust temperature. We have chosen 
to compute the color corrections on a per-MBB basis because this approach is 
very versatile; all possible two-component (and single-MBB) models are linear 
combinations of MBBs, so we can apply all of our color correction machinery 
even when we allow parameters other than temperature (e.g. $f_1$) to vary and 
thereby modify the dust spectrum shape.

With these color corrections and the  formalism established in 
$\S$\ref{sec:modeling} in hand, we can mathematically state the model we will 
use e.g. during MCMC sampling to predict the observed SED. The predicted 
observation in band $\nu_i$ is given by:

\begin{equation}
\label{eqn:inten}
\tilde{I}_{\nu_i} = \frac{\sum\limits_{k=1}^{2} b_{\nu_i}(T_k, \beta_k) f_k q_k (\nu_{i,c}/\nu_0)^{\beta_k} B_{\nu_{i,c}}(T_k) u_{\nu_i}^{-1}}{\sum\limits_{k=1}^{2} b_{545}(T_k, \beta_k) f_k q_k (545 \textrm{GHz}/\nu_0)^{\beta_k} B_{545}(T_k)}\tilde{I}_{545}
\end{equation}

This equation is quite similar to Equation \ref{equ:iras}, but with two 
important differences. First, the normalization of $\tilde{I}_{\nu_i}$ is now 
specified by $\tilde{I}_{545}$, which represents the IRAS convention 
\PLANCK~545 GHz intensity. The denominator serves to ensure that, for the case 
of $\nu_i$=545 GHz, $\tilde{I}_{545}$ is self-consistent. Second, each term in 
the numerator is multiplied by a unit conversion factor 
$u_{\nu_i}^{-1}$. This factor is necessary because some of the 
\PLANCK~maps of interest have units of $K_{CMB}$ (100-353 GHz), while the
remaining maps (545-3000 GHz) have units of MJy/sr. We have adopted the 
strategy of predicting each band in its native units, whether MJy/sr or 
$K_{CMB}$. For this reason, we always evaluate $B_{\nu_{i,c}}$ in Equation 
\ref{eqn:inten} in MJy/sr and let $u_{\nu_i}$=1 (dimensionless) for 
$\nu_i$$\ge$545 GHz. For $\nu_i$$\le$353 GHz, $u_{\nu_i}$ 
represents the conversion factor from $K_{CMB}$ to MJy/sr, given by 
\cite{planckresponse} Equation 32.

\section{Global Model Parameters}
\label{sec:global}

\begin{figure*}
\begin{center}
\epsfig{file=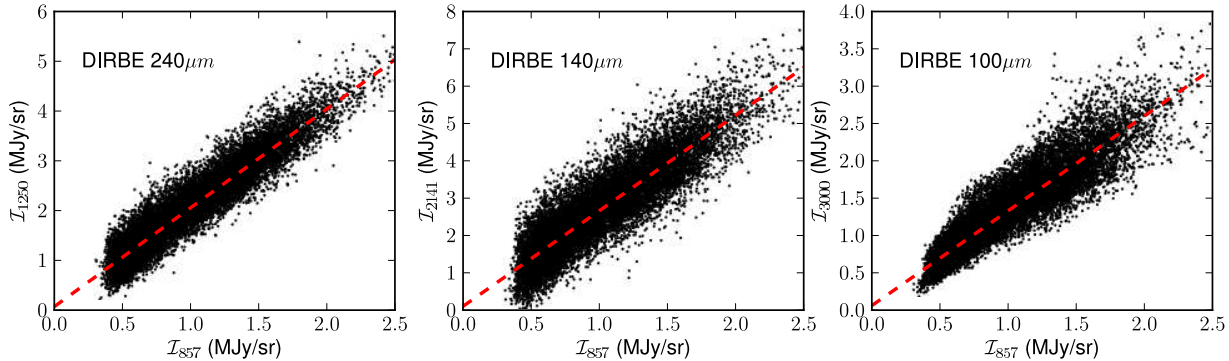, width=6.5in}
\caption{\label{fig:dirbe_slopes} Linear fits of SFD-reprocessed DIRBE 
240$\mu$m (left), 140$\mu$m (center), and 100$\mu$m (right) as a function of 
\PLANCK~857 GHz. The red lines illustrate the DIRBE correlation slopes used in 
our dust emission model optimization of $\S$\ref{sec:global}.}
\end{center}
\end{figure*}

While we ultimately aim to obtain \PLANCK-resolution maps of the spatially 
varying dust temperature and optical depth, we start by applying the 
machinery/formalism thus far developed to reassess the best-fit global 
two-component model parameters in light of the \PLANCK~HFI data.

FDS99 determined the best-fit values of the two-component model global 
parameters $\beta_1$, $\beta_2$, $q_1/q_2$ and $f_1$ via a correlation slope
analysis incorporating DIRBE and FIRAS data. Here we seek to estimate these 
same global parameters via an analogous correlation slope analysis in which we 
swap the \PLANCK~HFI maps for FIRAS at low frequencies, while still relying on 
DIRBE at higher frequencies. We also seek to determine via this correlation 
slope analysis whether or not the combination of \PLANCK+DIRBE data favors 
two-component models over single-MBB models in the same way that the 
FIRAS+DIRBE data did in the FDS99 analysis.

In the two-component model case, based on a spectrum of \PLANCK~and 
DIRBE correlations slopes, we wish to obtain estimates for six free parameters:
$\beta_1$, $\beta_2$, $q_1/q_2$, $f_1$, $T_2$ and the overall spectrum 
normalization $n$. The constraints we employ are the correlation slopes of each
of the \PLANCK~HFI bands, as well as DIRBE 100$\mu$m (3000 GHz), 140$\mu$m 
(2141 GHz) and 240$\mu$m (1250 GHz) relative to \PLANCK~857 GHz, i.e. 
$dI_{\nu_i}/dI_{857}$. We will refer to the slope for band $\nu_i$ relative to 
\PLANCK~857 GHz as $s_{857,\nu_i}$. The slopes for \PLANCK~100-545 GHz are taken 
to be those derived from the relative zero level fits of $\S$\ref{sec:relzero},
and are illustrated by the dashed red lines in the right-hand column plots of 
Figure \ref{fig:dip}. The 857 GHz slope is unity by definition.

At 1250, 2141 and 3000 GHz, we use the SFD-reprocessed DIRBE maps. For each
DIRBE band, we determine $s_{857,\nu_i}$ by performing a linear fit to DIRBE as a 
function of \PLANCK~857 GHz, after both have been zodiacal light subtracted and
smoothed to $1^{\circ}$ FWHM, also restricting to the low HI mask of 
$\S$\ref{sec:relzero} (see Figure \ref{fig:dirbe_slopes}).

Counting 857 GHz, we thus have nine correlation slope constraints for six 
free parameters. Including DIRBE 140$\mu$m and 240$\mu$m is critical in making 
the problem at hand sufficiently overconstrained, and also in providing 
information near the peak of the dust SED at $\sim$160$\mu$m, which is 
particularly sensitive to the presence of a single versus multiple MBB 
components. 

We assume an uncertainty of 2\% on each of the $s_{857,\nu_i}$ and minimize the 
chi-squared given by:

\begin{equation} \label{equ:chi2corr}
\chi^2 = \sum_{i=0}^{8}\frac{\big[s_{857,\nu_{i}}-n\frac{\tilde{I}_{\nu_i}(\beta_1, \beta_2, f_1, q_1/q_2, T_2)}{\tilde{I}_{857}(\beta_1, \beta_2, f_1, q_1/q_2, T_2)}\big]^2}{\sigma_{s_{857,\nu_i}}^2}
\end{equation}

Where $\nu_i$$\in$\{100, 143, 217, 353, 545, 857, 1250, 2141, 3000\} GHz. Note 
that this formula encompasses the general two-component case; in the single-MBB
case, we take $f_1$=0 and hence $q_1/q_2$, $\beta_1$ and $T_1$ are immaterial, 
but Equation \ref{equ:chi2corr} still applies. Note also that no `priors' are
included to preferentially drag our results towards agreement with those of 
FDS99. The correlation slopes $s_{857, \nu}$ and their adopted uncertainties
are listed in the fifth and sixth columns of Table \ref{table:offs}.

The results of our chi-squared minimization are listed in Table 
\ref{tab:global}. First (model 1), we fix $\beta_1$, $\beta_2$ ,$q_1/q_2$ and
$f_1$ to the best-fit values from the FDS99 analysis based on DIRBE+FIRAS. We
then allow $n$ and $T_2$ to vary so as to best match our 
DIRBE+\PLANCK~spectrum. This results in a reduced chi-squared of 
$\chi^2_{\nu}$=3.41. Reassuringly, $n$ is quite close to unity. It should be 
noted though that our best-fit $T_2$ is $\sim$0.5 K lower than that found by 
FDS99 for the same values of $\beta_1$, $\beta_2$, $q_1/q_2$ and $f_1$.

Next (model 2), we consider the fully-general two-component model, allowing all
six model parameters to vary. In this case, the reduced chi-squared of the best
fit parameters is $\chi^2_{\nu}=$1.33, signifying that our introduction of four
additional free parameters is justified. The best-fit $\beta_1$ and $\beta_2$ 
are both consistent with the corresponding FDS99 values to within 5\%. 
$q_1/q_2$=8.22 represents a $\sim$40\% lower value than found by FDS99, while
$f_1$=0.0458 represents a $\sim$25\% increase relative to FDS99. Again, our
best-fit high-latitude $T_2$ is $\sim$0.5 K lower than the typical value of 
$\langle T_2 \rangle$=16.2 K from FDS99.

Lastly, we calculate the optimal single-MBB fit to the \PLANCK+DIRBE 
correlation slope spectrum. The best-fit single MBB has $\beta$=1.59, 
$T$=19.63, and $\chi^2_{\nu}$=5.65, indicating a significantly worse fit to the
data than our best-fit two-component model (model 2). Thus, our \PLANCK+DIRBE 
correlation slope analysis has confirmed the main conclusion of FDS99 and 
others e.g. \cite{reach95}, that the FIR/submm dust SED prefers two MBBs to 
just one, but, for the first time, independent of FIRAS. Still, it is apparent 
that the improvement in $\chi^2_{\nu}$ for single-MBB versus double MBB models 
found here is substantially less dramatic ($\Delta\chi^2_{\nu}$=4.32) than that
found in FDS99 ($\Delta\chi^2_{\nu}$=29.2). This is likely attributable to the 
exquisite narrow-band frequency coverage of FIRAS, especially near the dust SED
peak, which makes FIRAS a better suited data set than \PLANCK~for a detailed 
analysis of the globally best-fit dust SED model. In $\S$\ref{sec:hier}, we
confirm the basic conclusions of this section via an approach in which we 
allow the dust temperature to vary spatially. The analysis of $\S$7.5 also
allows us to confirm the conclusions of this section while including a 
fully detailed uncertainty model; our assumption of 2\% per-band uncertainties 
on the correlation slopes is largely a statement that we seek a model
accurate to 2\% from 100-3000 GHz, although the fact that our $\chi^2_{\nu}$ 
values are order unity suggests that the assumed uncertainties are not grossly 
over or underestimated.

\section{MCMC Fitting Procedure}
\label{sec:fitting}

The following subsections detail our procedure for constraining the 
two-component dust emission model parameters which are permitted to vary
spatially. We use the MCMC procedure described to perform two types of fits: 
(1) full-resolution $6.1'$ fits, in which only  the SED normalization and dust 
temperatures vary spatially, and (2) lower-resolution fits in which $f_1$ is 
also allowed to vary from one line of sight to another.

\begin{figure}
\begin{flushleft}
\epsfig{file=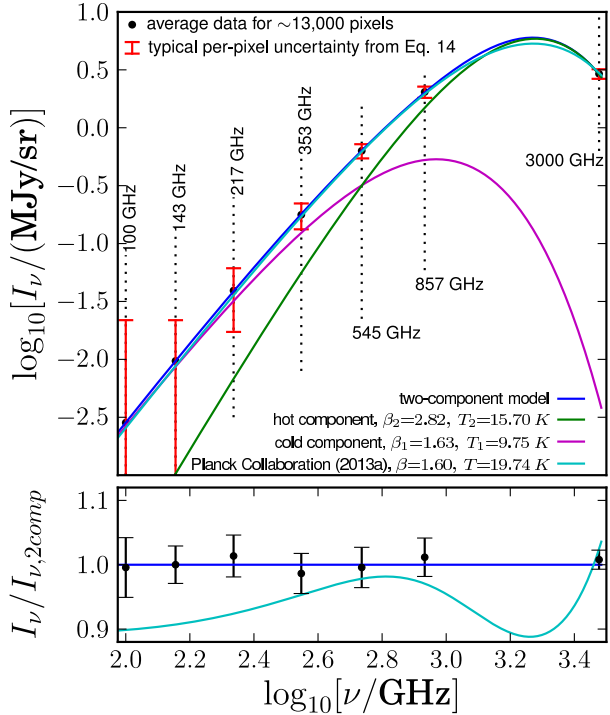, width=3.3in}
\caption{\label{fig:sed} Top: Summary of observed SEDs and best-fit thermal 
dust emission models for $\sim$13,000 $N_{side}$=2048 pixels with similar 
best-fit temperatures and optical depths (15.695 K$<$$T_2$$<$15.705 K, 
2.3$\times$$10^{-5}$$<$$\tau_{545}$$<$2.5$\times$$10^{-5}$). This region
of parameter space was arbitrarily chosen in order to obtain a large number of 
pixels within a narrow $T_2$ interval and small fractional range in 
$\tau_{545}$. Black points represent the average observed intensities after 
rescaling each pixel to $\tau_{545}$=2.4$\times$$10^{-5}$, while red error bars
represent the typical per-pixel uncertainties at each frequency. For each 
pixel, the best-fit two-component model is derived via the MCMC procedure of 
$\S$\ref{sec:mcmc}, based on \PLANCK~217-857 GHz and SFD 100$\mu$m at full 
$6.1'$ resolution. Note that the two lowest-frequency data points were not used
to derive the average two-component fit shown (blue line), while the three 
lowest-frequency data points were not used to derive the average 
\cite{planckdust} single-MBB fit shown (cyan line). Bottom: Comparison of 
average data, average two-component model and average \cite{planckdust} 
single-MBB model after dividing out the average two-component model. Black 
error bars represent the uncertainty on the mean observed spectrum. The 
two-component fit is consistent with the average data from 100-3000 GHz, 
whereas extrapolating the \cite{planckdust} model to 100-217 GHz yields 
predictions which are significantly low relative to the observed SED.}
\end{flushleft}
\end{figure}

\begin{figure*}
\begin{center}
\epsfig{file=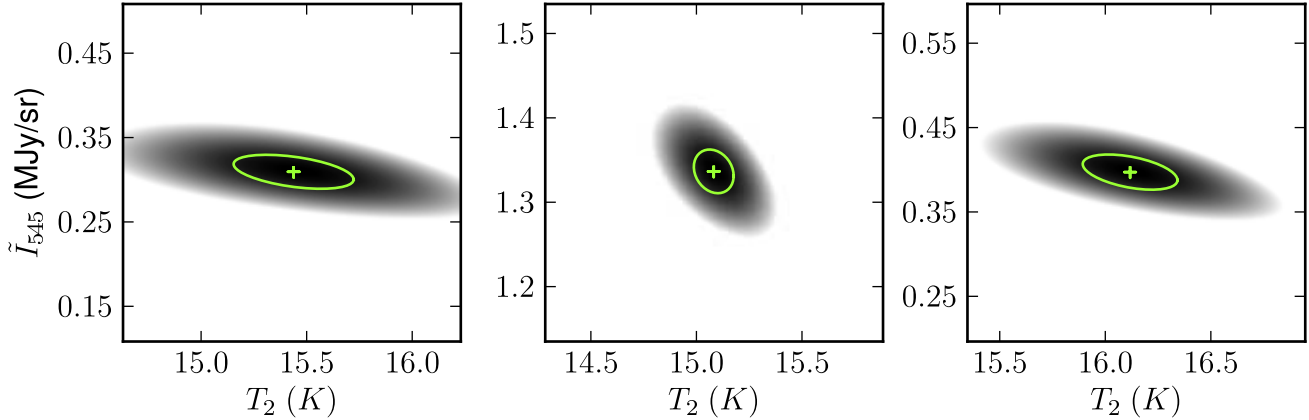, width=7.0in}
\caption{\label{fig:post} Gridded posterior PDFs for three $N_{side}$=2048 
HEALPix pixels, based on \PLANCK~217-857 GHz and SFD 100$\mu$m at full $6.1'$ 
resolution. The colorscale is linear in $\log(P)$, with black corresponding to 
the maximum of $\log(P)$ and white representing $max[\log(P)]-5$. Light green 
crosses and ellipses mark the best-fit parameters and $1\sigma$ uncertainties 
based on our MCMC sampling of the posteriors. Our MCMC parameter and 
uncertainty estimates are in good agreement with those based on gridded 
posteriors. These three pixels are also representative in that we find the 
posterior distributions from Equation \ref{eqn:post} are in general extremely 
well-behaved, showing no multimodality or other pathological qualities. Left: 
Low S/N pixel at high latitude in the Galactic north. Center: High S/N pixel in
the Polaris flare region. Right: Low S/N pixel at high latitude in the Galactic
south.}
\end{center}
\end{figure*}

\subsection{Pixelization}
\label{sec:pix}
For the purpose of fitting, we divide the sky into $\sim$50 million pixels of 
angular size $\sim$1.72$'$, defined by the HEALPix pixelization in Galactic 
coordinates, with $N_{side}$=2048. This pixelization is convenient because it 
is the format in which the \PLANCK~HFI maps were released, and because it 
adequately samples the $6.1'$ FWHM maps under consideration in our 
full-resolution fits. Our procedure will fit the intensity measurements in each
spatial pixel independently.

\subsection{Sampling Parameters}
\label{sec:samp}
As discussed in $\S$\ref{sec:modeling}, our full-resolution fits
consider the ``global'' parameters $f_1$, $q_1/q_2$, $\beta_1$, $\beta_2$ to be
spatially constant. We employ the best-fit \PLANCK+DIRBE global parameters of 
Table \ref{tab:global}, model 2. For each line of sight, only the dust spectrum
normalization and dust temperatures are allowed to vary. In order to predict 
the dust SED for a given pixel, we are thus left with two remaining degrees of 
freedom, and must choose an appropriate set of two parameters to sample and 
thereby constrain via MCMC. To determine the SED normalization in each pixel, 
we draw samples in $\tilde{I}_{545}$, the `IRAS convention' intensity in the 
\PLANCK~545 GHz bandpass, as defined in Equation \ref{eqn:inten}. With the four
aforementioned global parameters fixed, the dust spectrum shape is determined 
entirely by the two dust temperatures, which are coupled. To constrain the dust
temperatures, we sample in $T_2$, the hot dust temperature. For each sample in 
$T_2$, we compute the corresponding value of $T_1$, thereby fully specifying 
the SED shape. In principle, we could sample in either $T_1$ or $T_2$, but have
chosen to sample in $T_2$ because emission from this component dominates in the
relatively high frequency bands which most strongly constrain the dust 
temperatures.

For the lower resolution fits described in $\S$\ref{sec:lores}, we sample
in three parameters: $\tilde{I}_{545}$, $T_2$, and $f_1$.

\subsection{Markov Chains}
\label{sec:mcmc}

In our full-resolution fits, we use a MCMC approach to constrain the 
parameters $\tilde{I}_{545}$ and $T_2$. For each pixel, we run a 
Metropolis-Hastings (MH) Markov chain sampling the posterior probability of
the observed 217-3000 GHz thermal dust SED as a function of the two parameters 
$\tilde{I}_{545}$ and $T_2$. More specifically, for each pixel, we are sampling
the posterior given by:

\begin{equation}
\label{eqn:post}
P(\tilde{I}_{545}, T_2|\mathbf{I}) \propto \mathcal{L}(\mathbf{I}|\tilde{I}_{545}, T_2)P(T_2)P(\tilde{I}_{545})
\end{equation}

Here $\mathbf{I}$ denotes the vector of observed thermal dust intensities 
quoted under the `IRAS convention': $\mathbf{I}$ = ($I_{217}$, $I_{353}$,
$I_{545}$, $I_{857}$, $I_{3000}$). The likelihood function is given by:

\begin{equation} \label{equ:like}
\mathcal{L}(\mathbf{I}|\tilde{I}_{545}, T_2) = \textrm{exp}\Big[-\frac{1}{2}(\mathbf{I}-\mathbf{\tilde{I}})^T\Sigma^{-1}(\mathbf{I}-\mathbf{\tilde{I}})\Big]
\end{equation}

Here $\mathbf{\tilde{I}}$ is the vector of predicted observations 
based on Equation \ref{eqn:inten} and the proposed values of $\tilde{I}_{545}$ 
and $T_2$: $\mathbf{\tilde{I}} = $($\tilde{I}_{217}$, 
$\tilde{I}_{353}$, $\tilde{I}_{545}$, $\tilde{I}_{857}$, $\tilde{I}_{3000}$). 
$\Sigma$ is the per-pixel covariance matrix constructed based on the 
uncertainties in the observed intensities:

\begin{equation} \label{equ:covariance}
\Sigma = \begin{pmatrix} \sigma^2_{217}                           & \ldots  & \rho_{217,3000}\sigma_{217}\sigma_{3000} \\ 
                         \vdots                                   & \ddots  & \vdots                                   \\
                         \rho_{3000,217}\sigma_{3000}\sigma_{217} & \ldots  & \sigma^2_{3000}
\end{pmatrix}
\end{equation}

For each pixel $p$ in band $\nu_i$, the variance of the measured value 
$I_{\nu_i}(p)$ is taken to be:

\begin{multline} \label{equ:errorbudget}
\sigma^2_{\nu_i}(p) = c^2_{\nu_i}I_{\nu_i}^2(p) + c^2_{\nu_i}\sigma_{CMB, \nu_i}^2 + (\delta O_{\nu_i})^2 \\
+ n_{\nu_i}^2 + \sigma^2_{CO, \nu_i}(p)  + \sigma^2_{CIBA, \nu_i}
\end{multline}

This error budget is modeled after \cite{planckdust} Equation B.1, but with some 
modifications and additions. The first term accounts for the multiplicative 
uncertainty on the input maps. Table \ref{table:offs} lists the multiplicative 
calibration uncertainty $c_{\nu}$ for each band. These values are taken from 
Table 11 of \cite{planckcalib}. The second term represents an uncertainty due 
to our subtraction of the \verb|SMICA| CMB model. The analogous term in 
\cite{planckdust} Equation B.1 is ($c_{\nu}$$\times$\verb|SMICA|($p$))$^2$, i.e. 
an uncertainty proportional to the CMB model amplitude in each pixel. Because 
this term's spatial dependence can imprint the CMB anisotropies on the derived 
parameters, we have chosen to replace \verb|SMICA|($p$) with a spatially 
constant, RMS value for the CMB amplitude, $\sigma_{CMB, \nu_i}$. 
$\delta O_{\nu_i}$ represents the uncertainty in the band $\nu_i$ zero level 
offset, and the values of $\delta O_{\nu_i}$ can be read off from the second 
column of Table \ref{table:offs}. $n_{\nu_i}$ represents the instrumental noise
in band $\nu_i$. Because using per-pixel noise estimates based on the \PLANCK~
\verb|ii_cov| parameter can imprint features of the survey pattern onto the 
derived parameters, we have adopted a conservative, spatially constant value of
$n_{\nu_i}$ for each band. These values of $n_{\nu_i}$ are listed in Table 
\ref{table:offs}. The next term accounts for the uncertainty on the CO emission
correction, taking $\sigma_{CO, \nu_i}(p)$=0.15$\times$$I_{CO, \nu_i}(p)$ (see 
$\S$\ref{sec:mole}, specifically Equation \ref{equ:molcorr}).

Finally, we include a term to account for the RMS amplitude of the cosmic
infrared background anisotropy (CIBA) in band $\nu_i$, $\sigma_{CIBA, \nu_i}$. 
The values for the CIBA RMS amplitudes are obtained by assuming a $T$=18.3 K, 
$\beta$=1.0 MBB spectrum for the CIB, with 857 GHz normalization from 
\cite{ciba}. The CIBA not only contributes to the per-band variance 
$\sigma^2_{\nu_i}$, but also to the inter-frequency covariances; this is why we
have included the off-diagonal terms in the covariance matrix of Equation 
\ref{equ:covariance}. In our noise model, the CIBA is the only source of 
inter-frequency covariance. Thus, the off-diagonal covariance matrix element 
between bands $\nu_i$ and $\nu_j$ is given by:

\begin{equation}
\Sigma_{ij} = \rho_{\nu_i, \nu_j}\sigma_{\nu_i}\sigma_{\nu_j} = \rho_{CIBA, \nu_i, \nu_j}\sigma_{CIBA, \nu_i}\sigma_{CIBA, \nu_j}
\end{equation}

With values for $\rho_{CIBA, \nu_i, \nu_j}$ from \cite{covarciba}. The approach
we have taken in accounting for the CIBA is similar to that of 
\cite{planckdust}, Appendix C, in that we treat the CIBA amplitude in each 
pixel as a Gaussian random draw. However, instead of performing a separate 
analysis to gauge the uncertainty on derived dust parameters due to the CIBA, 
we allow the CIBA covariance to propagate naturally into our uncertainties via 
the likelihood function. Still, our treatment of the CIBA is a major 
oversimplification; a more sophisticated approach that accounts for the 
detailed CIBA spatial structure, or even removes the CIBA by subtraction would
be preferable. 

We include the following prior on the hot dust temperature:

\begin{equation} \label{equ:t2prior}
P(T_2) = \mathcal{N}(T_2|\bar{T}_2, \sigma_{\bar{T}_2})
\end{equation}

With $\bar{T}_2$ = 15.7 K and $\sigma_{\bar{T}_2}$ = 1.4 K. The $T_2$ prior 
mean is chosen based on the typical high-latitude $T_2$ value derived from
the correlation slope analysis of $\S$\ref{sec:global}. We find, as 
desired, that this relatively broad $T_2$ prior has little influence on the 
derived temperatures, other than to regularize the rare pixels with one or more
defective intensities which might otherwise yield unreasonable parameter
estimates. In principle, there can also be an informative prior on 
$\tilde{I}_{545}$. However, we have chosen to assume a uniform prior on the SED
normalization and, as a matter of notation, will omit $P(\tilde{I}_{545})$ 
henceforward. In practice we always perform computations using logarithms of 
the relevant probabilities.

\begin{figure}
\begin{center}
\epsfig{file=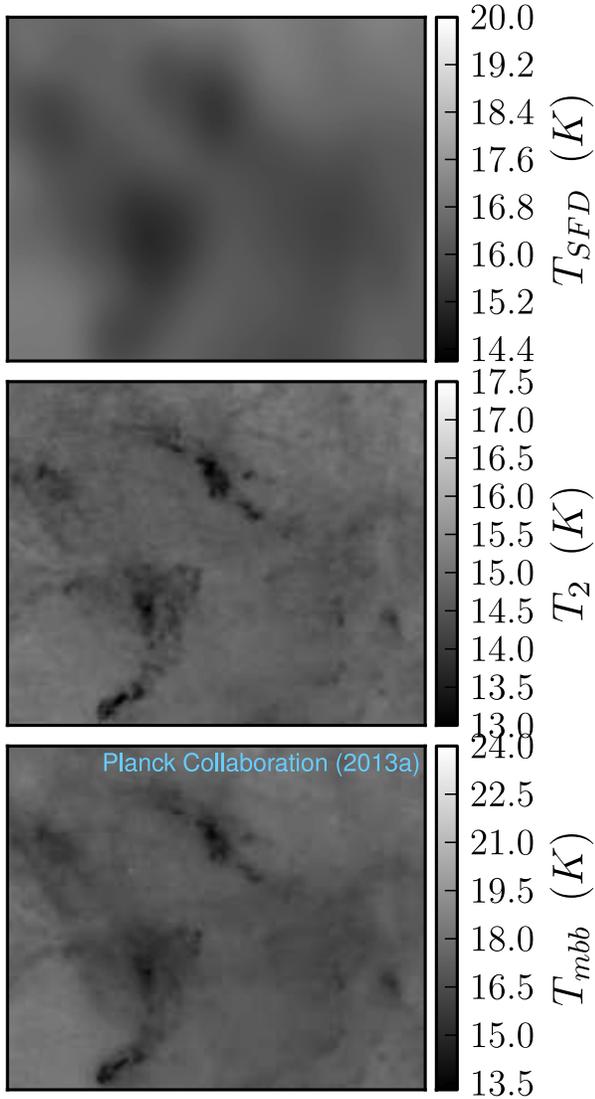, width=3.3in}
\caption{\label{fig:comparison} Comparison of temperature maps based on FIR 
dust emission over a $10.5^{\circ}\times8.3^{\circ}$ region centered about 
$(l,b) = (111.6^{\circ}, 18.3^{\circ})$. Top: SFD temperature map based on 
DIRBE $100\mu$m and $240\mu$m, with $\sim$1.3$^{\circ}$ resolution. Center: 
$6.1'$ resolution two-component temperature based on \PLANCK~217-857 GHz and 
SFD 100$\mu$m. Bottom: \cite{planckdust} single-MBB temperature map based on 
\PLANCK~353-857 GHz and 100$\mu$m data, with $5.1'$ FWHM. Both temperature maps
incorporating \PLANCK~observations clearly show a major improvement in angular 
resolution relative to SFD.}
\end{center}
\end{figure}

For each pixel, we initialize the Markov chain with parameters 
$\tilde{I}_{545}$ = $I_{545}$ and $T_2$ consistent with the FDS99 
DIRBE 100$\mu$m/240$\mu$m ratio map $\mathscr{R}$. The initial proposal 
distribution is a two-dimensional normal distribution, with 
$\sigma_{T_2}$=0.25 K, 
$\sigma_{\tilde{I}_{545}}$=max(0.01$\times$$I_{545}$, 0.05 MJy/sr) and
$\rho_{T_2, \tilde{I}_{545}}$=0. We run 5 iterations of burn-in, each 
consisting of 500 MH steps. After each burn-in iteration, we rescale the 
proposal distribution so as to ultimately attain an acceptance fraction 
$f_{acc}$ as close as possible to the optimal value $f_{opt}=0.234$. This is 
accomplished by multiplying the proposal distribution standard deviations by 
$f_{acc}$/$f_{opt}$.

After burn-in, we estimate the parameters and their uncertainties by performing
10,000 sampling steps, with $T_{2, j}$ and $\tilde{I}_{545, j}$ denoting the 
proposed parameter values at the $j^{th}$ step since the end of burn-in. From 
these 10,000 samples, we compute estimates of each parameter's mean, 
$\langle T_2 \rangle$ = $\langle T_{2, j} \rangle$, 
$\langle \tilde{I}_{545} \rangle$ = $\langle \tilde{I}_{545, j} \rangle$, of 
each parameter's variance, 
$\sigma^2_{T_2}$ = $\langle T^2_{2, j} \rangle-\langle T_{2, j} \rangle ^2$, 
$\sigma^2_{\tilde{I}_{545}}$=$\langle \tilde{I}^2_{545, j} 
\rangle-\langle \tilde{I}_{545, j} \rangle ^2$ and of the covariance 
$\sigma_{T_2}\sigma_{\tilde{I}_{545}}$=$\langle T_{2,j}-\langle T_2 \rangle 
\rangle \langle \tilde{I}_{545, j} - \langle \tilde{I}_{545} \rangle \rangle$. 

After obtaining this initial estimate of the covariance matrix for each pixel, 
we re-run a second iteration of the entire MCMC procedure, starting from the 
first burn-in period. On this iteration, for each pixel, we begin with a 
proposal distribution that is a two-dimensional Gaussian with covariance equal 
to the first-pass covariance estimate. This gives the each pixel's proposal 
distribution approximately the `right shape', whereas on the first pass we 
started by simply guessing the relative widths of the proposal distribution in 
$\tilde{I}_{545}$, $T_2$, and also assumed that the 
$\rho_{T_2, \tilde{I}_{545}}$=0. 

Lastly, during post burn-in sampling, we also estimate the monochromatic 
two-component intensity at 545 GHz, $M_{545}$ = $\langle M_{545, j} \rangle$ = 
$\langle \tilde{I}_{545, j}/b_{545}(T_{2,j}, \beta_2) \rangle$, its variance, 
and the 545 GHz optical depth $\tau_{545}$=$\langle \tau_{545, j} \rangle$ = 
$\langle M_{545, j}/S_{545, j} \rangle$ and its variance. $\tau_{545}$ and 
$M_{545}$ are more readily useful than the sampling parameters themselves for 
translating our fit results into predictions of reddening ($\S$\ref{sec:ebv}) 
and thermal dust emission ($\S$\ref{sec:lofreq}), respectively. At high 
Galactic latitude, we find a typical $T_2$ uncertainty of 0.45 K, and typical 
$\tilde{I}_{545}$ fractional uncertainty of 13\%. Figure \ref{fig:sed} 
illustrates the two-component model SED and the intensity measurements which 
constrain our fits, while Figure \ref{fig:post} shows example posterior PDFs 
for three pixels. Figure \ref{fig:comparison} shows a map of our derived hot 
dust temperature at full-resolution, for a patch of sky in the Polaris flare 
region.

We validated the parameters and uncertainties recovered from our MCMC procedure
by comparing with results based on finely gridded posterior calculations 
performed on a random subset of pixels. These comparisons verified that the 
proposal distribution rescaling and reshaping steps that we employ do improve 
the accuracy of the recovered parameters/uncertainties, and that the 
parameters/uncertainties ultimately derived are highly reliable. We can 
quantify the fidelity of our MCMC parameter estimates by noting that the RMS 
fractional discrepancy between MCMC and gridded posterior means is 0.25\% for
$\tilde{I}_{545}$ and 0.07\% ($\sim$0.01 K) for $T_2$. Regarding the accuracy 
of our uncertainty estimates, we find RMS fractional discrepancies of 2.2\% for
 $\sigma_{\tilde{I}_{545}}$ and 2.4\% for $\sigma_{T_2}$. Aside from these 
small statistical scatters, we find no biases in our MCMC estimates of the
 parameters and their uncertainties.

\subsection{Low-resolution Fits}
\label{sec:lores}

\begin{figure}
\begin{center}
\epsfig{file=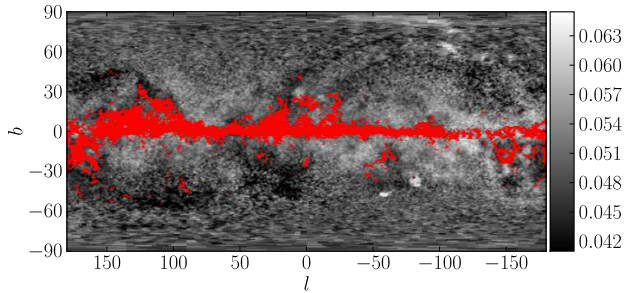, width=3.3in}
\caption{\label{fig:f1} $1^{\circ}$ FWHM full-sky map of $f_1$ derived from our
low-resolution fits described in $\S$\ref{sec:lores}. Red coloring masks pixels
with appreciable molecular emission, as defined in $\S$\ref{sec:mole}. Such 
pixels should not be trusted in this analysis, which is sensitive to the SED 
shape at low frequencies affected by CO line emission. Variations in $f_1$
along the ecliptic plane are spurious results of imperfect zodiacal light
subtractions. However, interesting astrophysical variations of $f_1$ 
are evident, particularly the trend of increasing $f_1$ with decreasing 
absolute Galactic latitude, the relatively low $f_1$ values in the Polaris 
flare and R Coronae Australis regions, and the clouds with relatively high 
$f_1$ values near the north Galactic pole.}
\end{center}
\end{figure}

As mentioned in $\S$\ref{sec:modeling}, the combination of high S/N and high 
angular resolution afforded by the \PLANCK~HFI maps provides us with the 
opportunity to allow additional parameters of the two-component model, 
previously fixed by FDS99, to vary spatially. Specifically, we consider
allowing $f_1$ to vary, while maintaining $\beta_1$, $\beta_2$, and $q_1/q_2$
spatially constant. In principle, we could alternatively introduce a third free
parameter by permitting $\beta_1$, $\beta_2$ or $q_1/q_2$ to vary while holding
$f_1$ fixed. However, a model in which $f_1$ varies continuously from one line 
of sight to another is the most natural three-parameter scenario, in that $f_1$
variation can be attributed to continuous changes in the dust species' mass
fractions, whereas continuous variations in the other global parameters, which 
represent grain emission/absoprtion properties, seem less plausible.

In order for our variable $f_1$ fits to remain sufficiently constrained 
following the introduction of a third free parameter, we enhance per-pixel S/N 
by smoothing the input maps to $1^{\circ}$ FWHM, and pixelize at $N_{side}=64$.
To best constrain the model parameters in each pixel, we also include 
\PLANCK~100 GHz and 143 GHz, and DIRBE 140$\mu$m and 240$\mu$m, all at 
$1^{\circ}$ resolution.

We now run Markov chains sampling in all three of $f_1$, $\tilde{I}_{545}$ and 
$T_2$, with the posterior given by:

\begin{equation}
\label{eqn:f1post}
P(\tilde{I}_{545}, T_2, f_1|\mathbf{I}) \propto \mathcal{L}(\mathbf{I}|\tilde{I}_{545}, T_2, f_1)P(T_2)P(f_1)
\end{equation}

The likelihood here is conceptually the same as that of Equation 
\ref{equ:like}, but now depends on $f_1$, which can vary from proposal to 
proposal within each individual pixel. The other difference is that 
$\mathbf{I}$ and $\mathbf{\tilde{I}}$ now include 100 GHz, 
143 GHz, 140$\mu$m and 240$\mu$m, in addition to the five bands used for the 
full-resolution fits.

The prior $P(T_2)$ from Equation \ref{equ:t2prior} 
remains unchanged. We adopt the following prior on $f_1$:

\begin{equation} \label{equ:f1prior}
P(f_1) = \mathcal{N}(f_1|\bar{f}_1, \sigma_{\bar{f}_1})
\end{equation}

With $\bar{f}_1$=0.0485 (from Table \ref{tab:global}, model 2) and 
$\sigma_{\bar{f}_1}$=0.005. This is a fairly stringent prior, but we must 
restrict the fit from wandering with too much freedom, as we are attempting to 
constrain three parameters using an SED with only nine intensity measurements, 
several of which are quite noisy. Again, we have adopted a uniform prior on 
$\tilde{I}_{545}$, and, as mentioned previously, we have omitted it from 
Equation \ref{eqn:f1post} as a matter of notation.

The resulting full-sky map of $f_1$ is shown in Figure \ref{fig:f1}. A 
general trend of increasing $f_1$ towards lower absolute Galactic latitudes is 
apparent. The other most salient features are the relatively low values of 
$f_1$ in the Polaris flare and R Coronae Australis regions, and the relatively 
high $f_1$ clouds near the north Galactic pole.

\subsection{Global Parameters Revisited}
\label{sec:hier}

The posterior sampling framework thus far described also affords us an
opportunity to evaluate the goodness-of-fit for competing dust SED models, and 
thereby cross-check the conclusions of our correlation slope analysis in
$\S$\ref{sec:global}. The basic idea will be to continue evaluating the 
posterior of Equation \ref{eqn:post}, but at low resolution ($N_{side}$=64), 
including all HFI bands as well as DIRBE 100$\mu$m, 140$\mu$m and 240$\mu$m, 
and switching to a uniform prior on $T_2$. Under these circumstances, the 
chi-squared corresponding to the best-fit parameters for pixel $p$, termed 
$\chi^2_p$, is simply $-2 \times\log(P_{max})$. We will refer to the
per-pixel chi-squared per degree of freedom as $\chi^2_{p, \nu}$.

\begin{figure}
\begin{center}
\epsfig{file=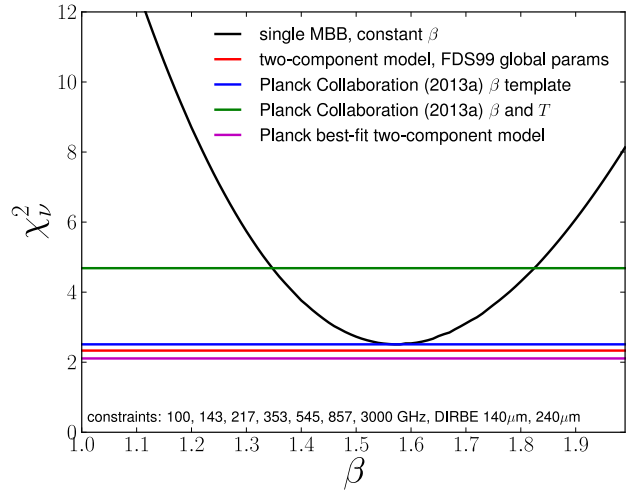, width=3.3in}
\caption{\label{fig:chi2_dirbe} Comparison of goodness-of-fit, 
$\chi^2_{\nu}$=$\langle \chi^2_{p, \nu} \rangle$, for various dust SED models,
as described in $\S$\ref{sec:hier}. For single-MBB models with spatially 
constant $\beta$, we varied $\beta$ between 1 and 2 (horizontal axis), 
achieving reduced chi-squared $\chi^2_{\nu}$ shown by the black line, with 
$\beta$=1.57 providing the best single-MBB fit. Horizontal lines indicate 
$\chi^2_{\nu}$ for other dust emission models considered, including the FDS99 
best-fit two-component model (Table \ref{tab:global}, model 1, red) and the 
\cite{planckdust} single-MBB model (green). The minimum $\chi^2_{\nu}$ is 
achieved with two-componenent `model 2' from Table \ref{tab:global} (magenta).}
\end{center}
\end{figure}

Because we seek to compare the goodness-of-fit for various dust SED models in 
the diffuse ISM, we restrict to a set of $\sim$10,800 pixels ($\sim$22\% of the
sky), with $|b|>30^{\circ}$ and $|\beta|>10^{\circ}$. We also avoid the SMICA 
inpainting mask, pixels with appreciable CO contamination, and compact sources.
The goodness-of-fit `objective function' we employ to judge the quality of 
a particular dust SED model is $\langle \chi^2_{p, \nu} \rangle$, where the 
average is taken over the aforementioned set of $\sim$10,800 pixels. 
$\langle \chi_{p, \nu}^2 \rangle$ is also equivalent to the reduced 
chi-squared, $\chi^2_{\nu}$, when considering the total number of free 
parameters to be the number of pixels multiplied by the number of free 
parameters per pixel (and similarly for the total number of constraints), and
taking $\chi^2$=$\sum\chi^2_{p}$.

We calculate $\chi^2_{\nu}$ for various dust SED models, independently 
minimizing each $\chi^2_p$ by finding pixel $p$'s best-fitting dust 
temperature and normalization, then evaluating 
$\langle \chi_{p, \nu}^2 \rangle$. First, we consider single-MBB models with 
$\beta$ spatially constant (see the black line in Figure \ref{fig:chi2_dirbe}).
$\beta$=1.57 yields the best fit, with $\chi^2_{\nu}$=2.51. This result is in 
excellent agreement with that of $\S$\ref{sec:global}, where we found the 
best-fit single-MBB model to have $\beta$=1.59. 

We also evaluated $\chi^2_{\nu}$ for single-MBB models in which $\beta$ varies 
spatially. In these cases, we adopted the 0.5$^{\circ}$ resolution $\beta$ map 
from \cite{planckdust}. We started by calculating $\chi^2_{\nu}$ using the 
\cite{planckdust} temperature map, finding $\chi^2_{\nu}$=4.68. Note that in 
this case no per-pixel chi-squared minimization was involved, as we simply 
evaluated $\chi^2_{p}$ for each pixel based on the fully-specified 
\cite{planckdust} emission model. Next, we tested a single-MBB model for which 
we adopted the \cite{planckdust} $\beta$ map, but allowed the per-pixel 
temperature and normalization to vary so as to minimize $\chi^2_p$. In this 
case, we found $\chi^2_{\nu}$=2.51, effectively identical to the value found 
for the spatially constant $\beta$=1.57 single-MBB model. This is perhaps 
unsurprising, as the average $\beta$ value from \cite{planckdust} over the mask
in question is $\langle\beta\rangle$=1.58. This result does suggest, however, 
that in diffuse regions the half-degree variations in $\beta$ are not 
materially improving the goodness-of-fit over the full frequency range 100-3000
GHz relative to a model with appropriately chosen spatially constant $\beta$.

We move on to evaluate two-component models, first calculating $\chi^2_{\nu}$ 
with the FDS99 global parameters (Table \ref{tab:global}, model 1). We find
$\chi^2_{\nu}$=2.33, a slight improvement relative to the best-fitting
single-MBB models. Finally, we calculate $\chi^2_{\nu}$ for Table 
\ref{tab:global} model 2, the two-component model favored by our \PLANCK+DIRBE 
correlation slopes. In this case, we achieve the best goodness-of-fit out of 
all the models we have tested, with $\chi^2_{\nu}$=2.11.

Thus, our degree-resolution goodness-of-fit analysis has generally confirmed
the conclusions of $\S$\ref{sec:global}. We find the single-MBB $\beta$ value 
favored by the combination of \PLANCK~and DIRBE to be nearly identical here 
($\beta$=1.57) versus in $\S$\ref{sec:global} ($\beta$=1.59). As in 
$\S$\ref{sec:global}, we also find that the \PLANCK+FIRAS and \PLANCK+DIRBE 
best-fit two-component models from Table \ref{tab:global} outperform single-MBB
alternatives, though only by a relatively small margin in $\chi^2_{\nu}$. 
Still, because our present analysis has $\sim$75,500 degrees-of-freedom, 
$\Delta \chi^2_{\nu}$=0.4 formally corresponds to an enormously significant
improvement in $\chi^2$. The agreement between our correlation slope analysis 
and the present goodness-of-fit analysis is especially encouraging for three 
main reasons: (1) in the present analysis, dust temperature has been allowed to
vary on degree scales, whereas in $\S$\ref{sec:global} we assumed a single 
global dust temperature (2) the present analysis employs a fully detailed, 
per-pixel uncertainty model and (3) in the present analysis, our zero-level 
offsets factor into the dust temperature, whereas in $\S$6 this was not the 
case, meaning the former and latter analyses agree in spite of their potential 
to be affected by rather different systematics.

\section{Optical Reddening}
\label{sec:ebv}

While the temperature and optical depth maps thus far derived are useful for 
making thermal dust emission foreground predictions, estimating optical 
reddening/extinction is another important application of the $\tau_{545}$ map. 
Translating our two-component optical depth to reddening is especially valuable
because our $T_2$ map has $\sim$13$\times$ better angular resolution than the 
SFD temperature correction, and thus there is reason to believe our 
two-component reddening estimates may be superior to those of SFD. However, as 
discussed in $\S$\ref{sec:replace}, we do not yet advocate for the wholesale 
replacement of SFD, and more detailed work is still necessary to 
determine/quantify the extent to which \PLANCK-based dust maps might improve 
reddening estimates relative to SFD.

\subsection{Reddening Calibration Procedure}
\label{sec:calib_ebv}

\begin{figure}
\begin{center}
\epsfig{file=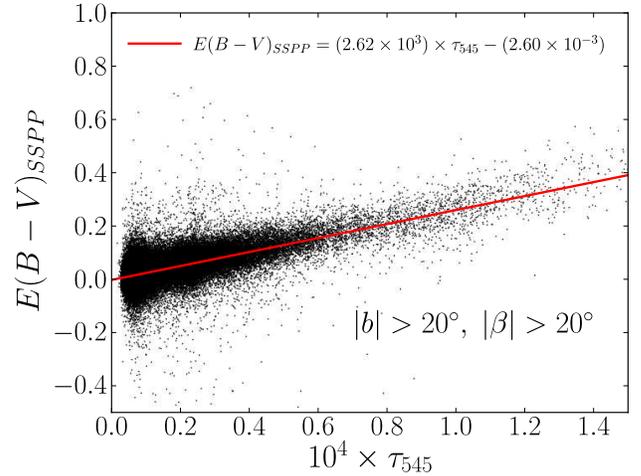, width=3.3in}
\caption{\label{fig:calib} Linear fit of $E(B-V)_{SSPP}$ as a function of
two-component 545 GHz optical depth, illustrating our procedure for 
calibrating optical depth to reddening, as described in 
$\S$\ref{sec:calib_ebv}.}
\end{center}
\end{figure}

We calibrate optical depth to reddening empirically rather than derive a 
relationship between $\tau_{545}$ and reddening by introducing additional 
assumptions about the dust grain physics and size distribution. To achieve this
empirical calibration, we must adopt a set of calibrator objects for which true
optical reddening is known. There are various possibilities at our disposal. 
\cite{planckdust} calibrated their radiance and $\tau_{353}$ maps to $E(B-V)$ 
using broadband Sloan Digital Sky Survey \citep[SDSS;][]{sdss} photometry for a
set of $\sim$10$^5$ quasars. The SFD calibration was originally tied to a 
sample of 384 elliptical galaxies, but was later revised by 
\citet[hereafter SF11]{schlafly11} based on $\sim$260,000 stars with both 
spectroscopy and broadband photometry available from the SEGUE Stellar 
Parameter Pipeline \citep[SSPP,][]{sspp}.

\begin{figure*}
\begin{center}
\epsfig{file=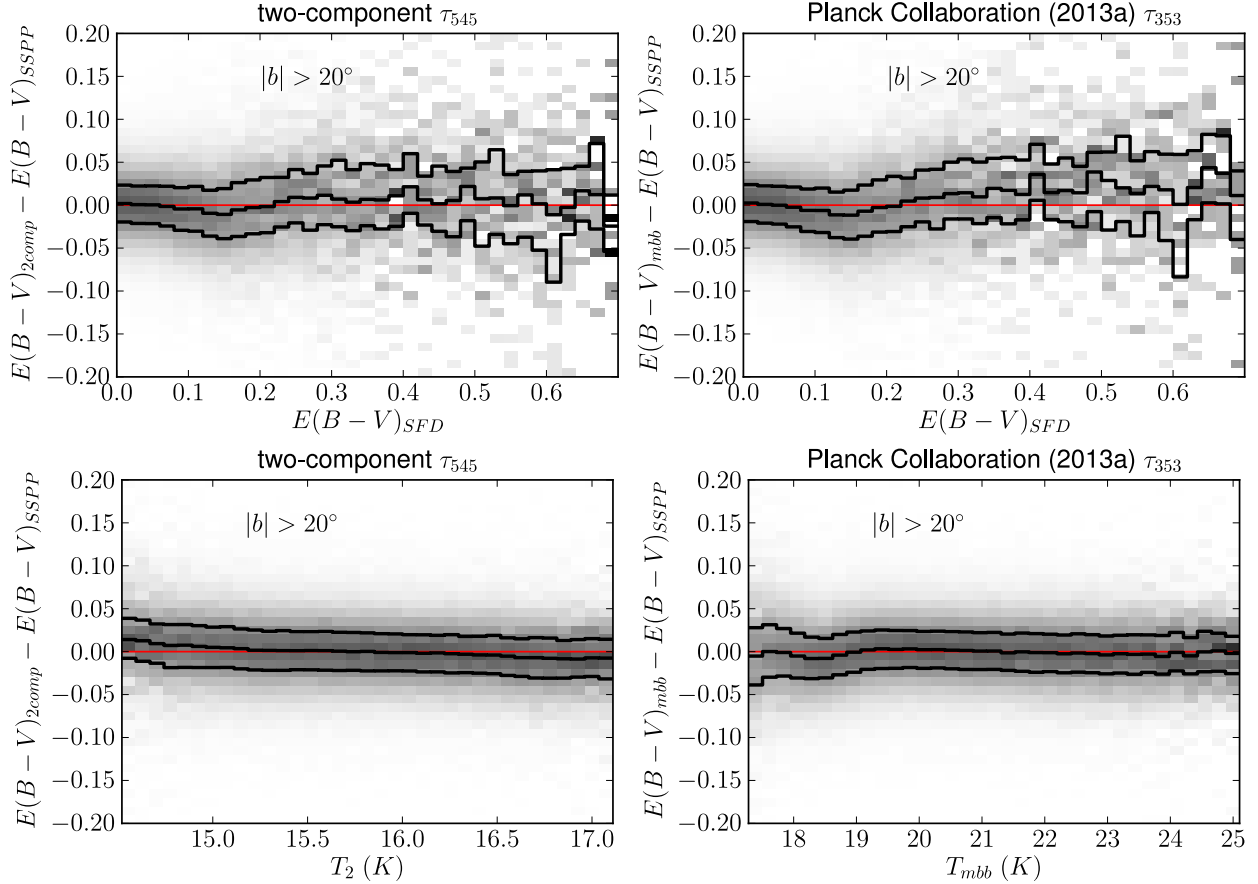, width=6.6in}
\caption{\label{fig:resid} (top left) Residuals of $E(B-V)_{2comp}$ relative to
$E(B-V)_{SSPP}$ as a function of $E(B-V)_{SFD}$. The grayscale represents the 
conditional probability within each $E(B-V)_{SFD}$ bin. The central black line 
shows the moving median. The upper and lower black lines represent the moving 
75th and 25th percentiles respectively. (bottom left) Residuals of 
$E(B-V)_{2comp}$ relative to $E(B-V)_{SSPP}$ as a function of hot dust 
temperature $T_2$. (top right) Same as top left, but illustrating the residuals
of $E(B-V)_{mbb}$, our calibration of the \cite{planckdust} $\tau_{353}$ to 
$E(B-V)_{SSPP}$. (bottom right)  Same as bottom left, but showing the 
$E(B-V)_{mbb}$ residuals as a function of the single-MBB dust temperature from 
\cite{planckdust}. The temperature axes always range from the 0.4$^{th}$ 
percentile temperature value to the 99.6$^{th}$ percentile temperature value.}
\end{center}
\end{figure*}

To calibrate our two-component optical depth to reddening, we make use of the 
stellar sample from SF11. Given a library of model stellar atmospheres, the 
spectral lines of these stars can be used to predict their intrinsic optical 
broadband colors. The `true' reddening is then simply the difference between 
the observed $g-r$ color and the $g-r$ color predicted from the spectral lines.
Applying a color transformation then yields `true' $E(B-V)$ values for 
$\sim$260,000 lines of sight. Throughout our SSPP calibration analysis, we 
restrict to the $\sim$230,000 lines of sight with $|b|$$>$20$^{\circ}$ in order
to avoid stars which may not lie behind the full dust column. In this section 
and $\S$8.2, we make absolute latitude cuts (in both $b$ and $\beta$) at 
$20^{\circ}$, to match the footprint of SF11 and adapt to the non-uniform 
distribution of SSPP stars on the sky. The calibration of two-component optical
depth to $E(B-V)$ is performed as a linear regression of $E(B-V)_{SSPP}$ versus
$\tau_{545}$. $\tau_{545}$ is considered to be the independent variable in this
regression, as we ultimately wish to predict $E(B-V)$ as a function of optical 
depth, and $\tau_{545}$ has much higher S/N than the SSPP $E(B-V)$ estimates.

This regression is illustrated in Figure \ref{fig:calib}. As expected, there 
is a strong linear correlation between $E(B-V)_{SSPP}$ and $\tau_{545}$. The 
conversion factor from $\tau_{545}$ to $E(B-V)$ is 2.62$\times$10$^{3}$. 
Reassuringly, the best-fit offset is close to zero, $\sim$2.6 mmag.

Figure \ref{fig:resid} shows the residuals of our $\tau_{545}$-based reddening 
predictions, $E(B-V)_{2comp}$, relative to the corresponding SF11 reddening 
measurements, $E(B-V)_{SSPP}$, as a function of SFD reddening, $E(B-V)_{SFD}$, 
(top left panel) and as a function of hot dust temperature (bottom left panel).
For comparison, the right panels show analogous residual plots, but with 
respect to reddening predictions based on our calibration of the 
\cite{planckdust} 353 GHz optical depth to $E(B-V)_{SSPP}$ , using the same 
regression procedure employed to calibrate $E(B-V)_{2comp}$. We refer to these 
reddening predictions based on the \cite{planckdust} single-MBB model and 
calibrated to the SF11 measurements as $E(B-V)_{mbb}$.

All four residual plots in Figure 12 show systematic problems at some level. 
The most striking systematic trend is the `bending' behavior of the reddening
residuals versus $E(B-V)_{SFD}$ (top panels), with the median residual 
bottoming out near $-10$ mmag at $E(B-V)_{SFD}$$\approx$0.15 mag. This behavior
is common to both $E(B-V)_{2comp}$ and $E(B-V)_{mbb}$, and in fact was first 
noted in the residuals of $E(B-V)_{SFD}$ itself relative to $E(B-V)_{SSPP}$ by 
SF11 (see their Figure 6). Such a bending behavior is troubling because it 
could indicate a nonlinearity common to many FIR reddening predictions based on
column densities inferred from dust emission. Alternatively, because the SF11 
stars are distributed over the sky in a highly non-uniform manner, the bend 
could arise from aliasing of discrepancies particular to certain sky regions 
(e.g. inner vs. outer Galaxy) on to the $E(B-V)_{SFD}$ axis.

\begin{figure*}
\begin{center}
\epsfig{file=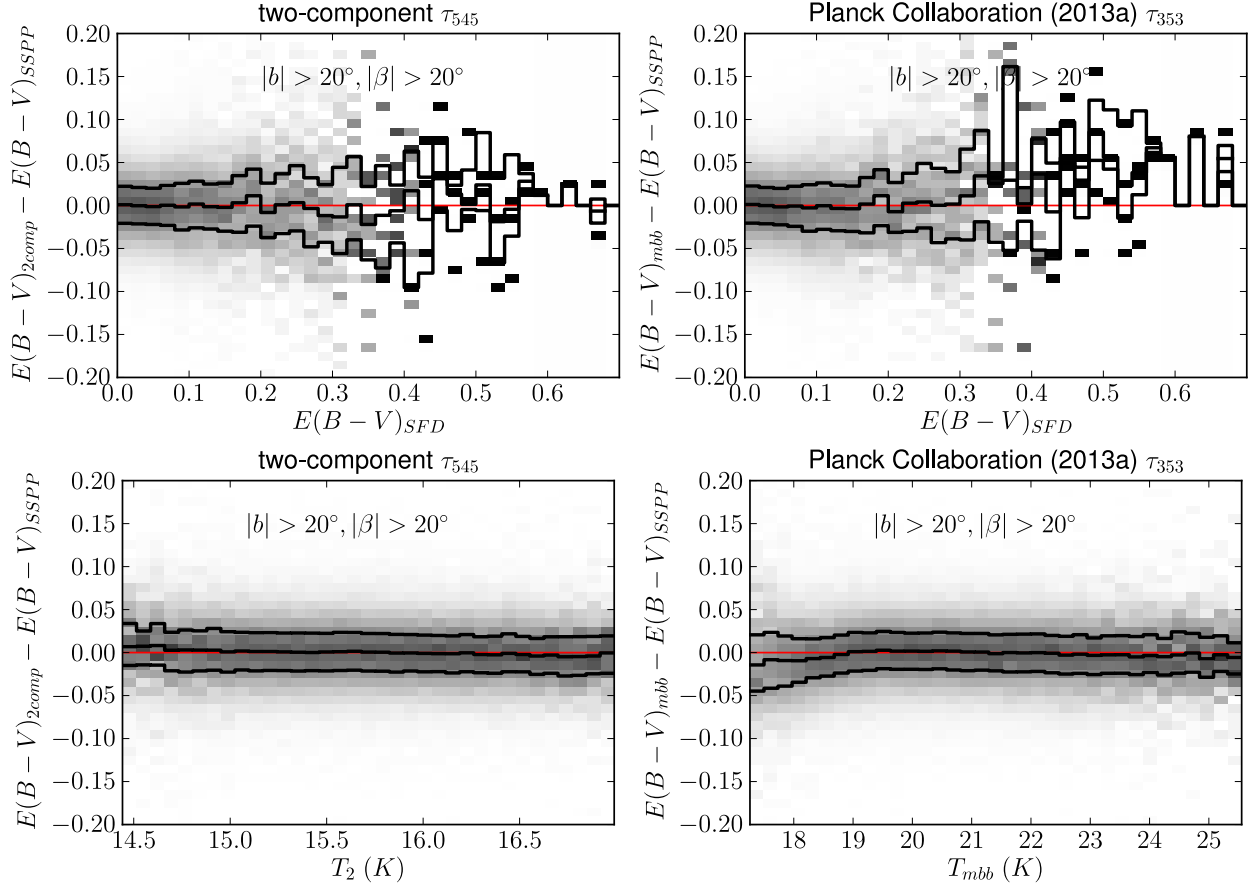, width=6.6in}
\caption{\label{fig:resid_ecl} Same as Figure \ref{fig:resid}, but restricting 
to high ecliptic latitude, $|\beta|>20^{\circ}$. In both the top left and top 
right plots, the bending of the reddening residuals as a function of 
$E(B-V)_{SFD}$ seen in Figure \ref{fig:resid} has been eliminated. Further, the
two-component reddening residual temperature dependence (bottom left) has been 
significantly reduced relative to the corresponding trend shown in Figure 
\ref{fig:resid}. For $E(B-V)_{SFD}$$\gtrsim$0.3 mag, the top row plots appear 
noisy because there are an insufficient number of remaining SSPP points of 
comparison.}
\end{center}
\end{figure*}

The obvious culprit for any potential nonlinearity in FIR-based reddening
estimates is a faulty temperature correction. For this reason, we have included
the bottom panels of Figure \ref{fig:resid}, to check for the presence of a 
temperature dependence of the reddening residuals. Indeed, in both the
two-component and single-MBB cases there exists some systematic dependence of
the reddening residuals on temperature. For $T_{mbb}$$\gtrsim$19 K, the median 
residual is reasonably flat, but at lower temperatures (the lowest temperature 
$\sim$20\% of SSPP sight lines), the median shows trends at the $\sim$10 mmag 
level. On the other hand, the median residual in the two-component case trends 
downward with increasing $T_2$ over the entire $T_2$ range shown, with a 
peak-to-peak amplitude of $\sim$20 mmag.

\subsection{Rectifying the Reddening Residuals}

In this section we describe our attempts to eliminate the systematic problems 
in the two-component reddening residuals shown in the left column of Figure 
\ref{fig:resid}. We employed two main strategies: (1) recomputing the 
two-component $\tau_{545}$ by re-running our Markov chains after modifying the 
input maps and/or changing the particular two-component model paramters adopted
and (2) making spatial cuts to isolate sky regions in which the residuals are 
especially pristine (or especially problematic).

The following is a list of dust model modifications we tested, but which proved
to have little impact on the reddening residual trends as a function of either 
$E(B-V)_{SFD}$ or $T_2$:

\begin{itemize}
\item Varying each of the global two-component model parameters $\beta_1$, 
$\beta_2$, $q_1/q_2$ and $f_1$ individually while holding the others fixed.
\vspace{-3mm}
\item Allowing $f_1$ to vary spatially as in the fits of $\S$\ref{sec:lores}.
\vspace{-2mm}
\item Changing the mean and/or variance of the $T_2$ prior.
\vspace{-6mm}
\item Varying multiple global parameters at a time e.g. both $f_1$ and 
$q_1/q_2$, restricting to regions of parameter space favored by our 
goodness-of-fit analyses described in $\S$\ref{sec:global} and 
$\S$\ref{sec:hier}.
\end{itemize}

We additionally investigated the following spatial cuts which did not resolve 
the dominant problems noted in the reddening residuals:
\begin{itemize}
\item Separating Celestial north and south.
\vspace{-3mm}
\item Separating Galactic north and south.
\vspace{-3mm}
\item Separating inner and outer Galaxy.
\vspace{-3mm}
\item Combining the above two sets of cuts i.e. separating the Galaxy into
quadrants. 
\vspace{-3mm}
\item Combining these spatial cuts with the dust model changes of the previous
list.
\end{itemize}

However, we found that changing the zero level offsets of the input maps 
had a significant effect on the strength of the anticorrelation between median 
reddening residual and $T_2$. In particular, we experimented with perturbing 
the zero level offset of \PLANCK~857 GHz while correspondingly changing the 
zero levels of the remaining \PLANCK~maps based on the prescription of 
$\S$\ref{sec:relzero}. We also experimented with changing the zero level of SFD
\verb|i100|, independent of the other zero levels. Unfortunately, completely
flattening the reddening residual dependence on $T_2$ required unreasonably 
large zero level modifications. For example, flattening the $T_2$ residual 
required adding $\gtrsim$0.6 MJy/sr to the \verb|i100| map. Such an offset is 
implausible, being an order of magnitude larger than the nominal \verb|i100| 
zero level uncertainty quoted by SFD, and comparable to the entire 3000 GHz CIB
monopole signal. Furthermore, we note that even these large zero level 
modifications had virtually no effect in eliminating the reddening residual 
`bend' versus $E(B-V)_{SFD}$. Thus, changing the zero level offsets showed 
hints of promise in rectifying the reddening residual temperature dependence, 
but could not by itself completely resolve the systematic trends in reddening 
residuals.

The only solution we have been able to identify that both removes the `bend' 
vs. $E(B-V)_{SFD}$ and simultaneously reduces the temperature dependence of the
reddening residuals is cutting out the ecliptic plane by restricting to 
$|\beta|>20^{\circ}$. In this case, we completely eliminated the bending 
behavior of the residual versus $E(B-V)_{SFD}$, and significantly reduced the 
$T_2$ dependence to a peak-to-peak amplitude of only $\sim$10 mmag (see Figure 
\ref{fig:resid_ecl}). Figure \ref{fig:resid_ecl} still includes the single-MBB 
plots (right column), to show that the bend versus $E(B-V)_{SFD}$ is eliminated
by the $|\beta|$ cut, even for the single-MBB model. However, the single-MBB 
residuals still differ systematically from zero for $T$$\lesssim$$19$ K. 
Perhaps the improvements in the two-component reddening residuals after 
restricting to high ecliptic latitude should come as no surprise, given that 
the ecliptic plane is the most obvious systematic problem with our temperature 
map (see the full-sky results shown in Figure \ref{fig:results}).

\begin{figure}
\begin{center}
\epsfig{file=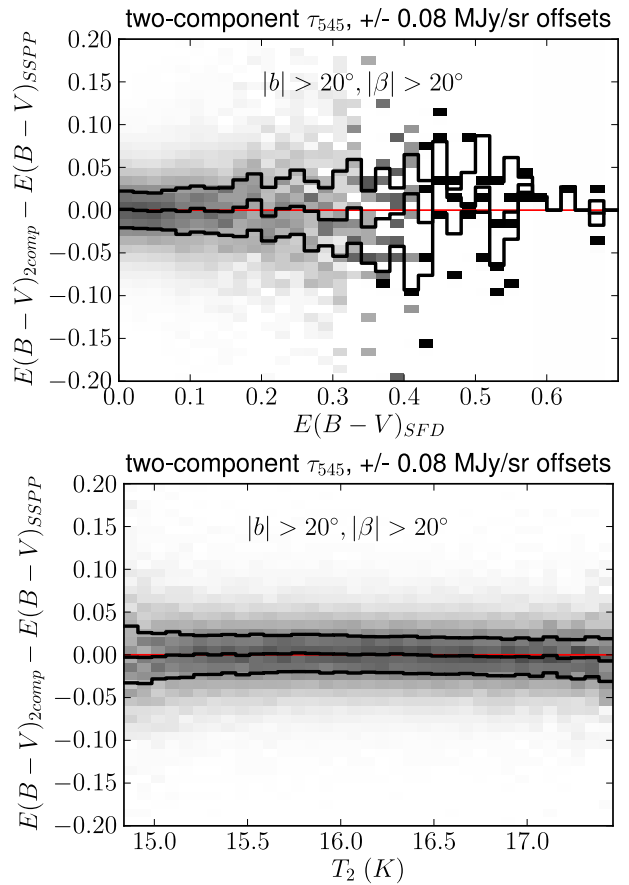, width=3.3in}
\caption{\label{fig:resid_offs} Two-component reddening residuals after 
restricting to high ecliptic latitude ($|\beta|>20^{\circ}$) and perturbing the
\texttt{i100} and 857 GHz zero levels by $+$0.08 MJy/sr and $-$0.08 MJy/sr 
respectively. The bending behavior as a function of $E(B-V)_{SFD}$ has been
eliminated, and virtually no temperature dependence remains. For 
$E(B-V)_{SFD}$$\gtrsim$0.3 mag, the top plot appears noisy because there are an
insufficient number of remaining SSPP points of comparison following our cut on
ecliptic latitude.}
\end{center}
\end{figure}

After cutting the ecliptic plane, we found that only small zero level
perturbations were required to fully flatten the temperature residuals, while 
still maintaining flat residuals versus $E(B-V)_{SFD}$. The optimal offsets we 
found were $\pm$0.08 MJy/sr to \verb|i100| and 857 GHz respectively (see Figure
\ref{fig:resid_offs}). These offsets are well within reason, given the nominal 
zero level uncertainties quoted in Table \ref{table:offs}.

\section{Comparison of Emission Predictions}
\label{sec:em_compare}

\subsection{The 353-3000 GHz Frequency Range}
\label{sec:hifreq}
Here we compare our two-component emission predictions to those of the 
\cite{planckdust} single-MBB model in the 353-3000 GHz range. This frequency 
range represents the overlap between the recommended range of applicability for
the \cite{planckdust} model and the 100-3000 GHz frequency range of our 
two-component model. Since we have used input maps that are very similar to 
those of \cite{planckdust}, and since our model and the \cite{planckdust} model
both fit the data well in this frequency range, good agreement between our 
two-component predictions and those of the \cite{planckdust} single-MBB model 
is to be expected.

\begin{figure} [ht]
\begin{center}
\epsfig{file=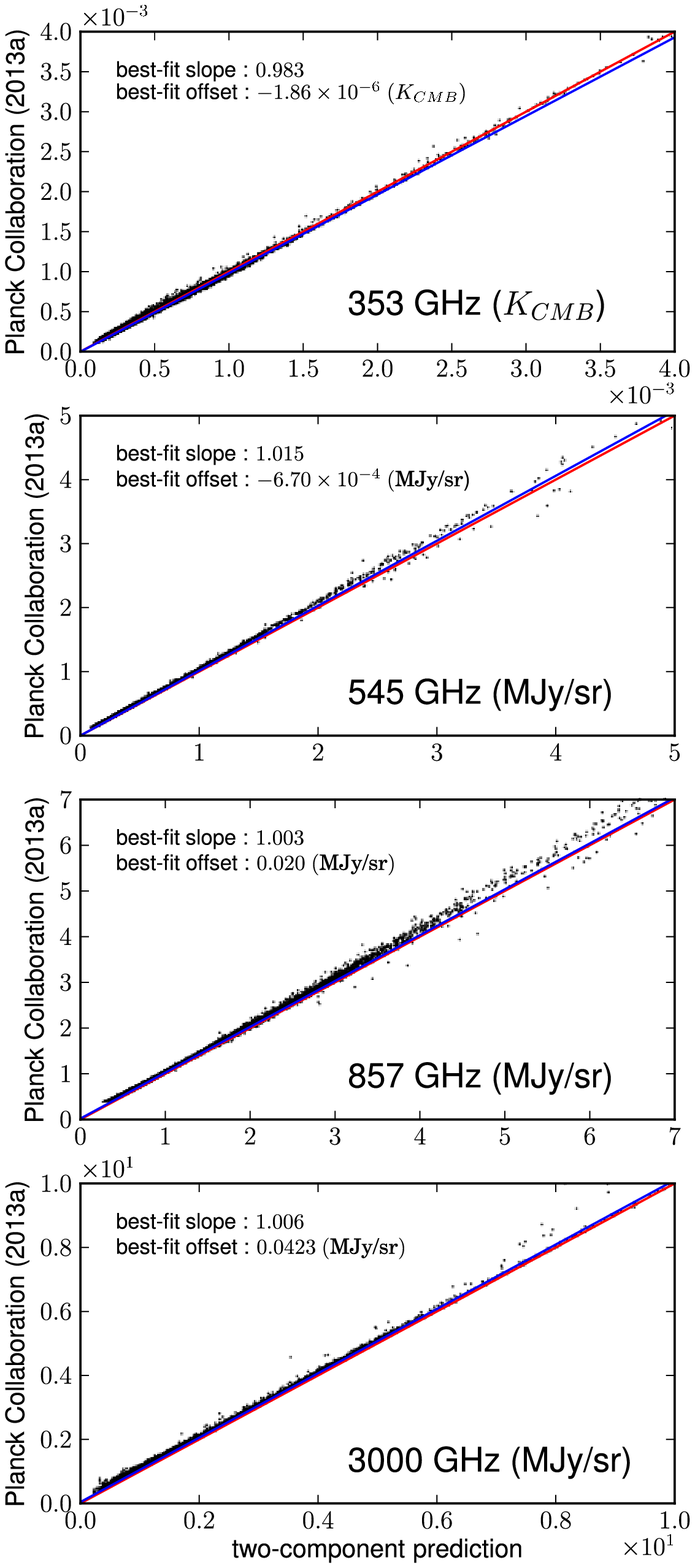, width=3.2in}
\caption{\label{fig:compare_hifreq} Scatter plots of \cite{planckdust} 
single-MBB predictions (vertical axes) versus our two-component predictions
(horizontal axes), rebinning to $N_{side}$=64 and restricting to the diffuse 
regions of $\S$\ref{sec:hier}. The lines of best fit are shown in blue, and red
lines represent perfect agreement between the two predictions. Note that a 
per-band offset has been applied to the \cite{planckdust} predictions to
account for the differing zero level offsets used in building the two models. 
After accounting for the different zero levels, the best fit offsets between 
predictions are consistent with zero to within the uncertainties quoted in 
Table \ref{table:offs}. The slopes are also very nearly unity, to within 
$\leq$1.7\%.}
\end{center}
\end{figure}

We compare the emission models in this frequency range by using each model in 
turn to predict the observed \PLANCK~353, 545, and 857 GHz maps, as well as the
3000 GHz DIRBE/\IRAS~map. We rebin to $N_{side}$=64 and restrict to the diffuse
sky regions of our mask from $\S$\ref{sec:hier}. We summarize this comparison 
by producing a per-band scatter plot of the \cite{planckdust} prediction versus
the two-component prediction, and performing a linear regression between these 
two quantities. Before plotting and performing these regressions, we adjusted 
the \cite{planckdust} predictions to account for the differing zero level 
offsets used in this work and in \cite{planckdust}. For instance, at 3000 GHz, 
\cite{planckdust} added 0.17 MJy/sr to the SFD98 zero level, whereas we made no
such modification; therefore, for the sake of comparison, we subtracted 0.17 
MJy/sr from the \cite{planckdust} predictions before plotting and performing 
the 3000 GHz regression.

The slopes obtained from these linear fits indicate very good agreement between
the single-MBB and two-component models, with values between 0.983-1.015 
(agreement at the $\leq$1.7\% level). The offsets are also consistent with zero
to within the uncertainties quoted in Table \ref{table:offs}. We do not find 
evidence that our two-component model provides emission predictions in the 
353-3000 GHz range which are superior to those of \cite{planckdust}. From 
353-3000 GHz and in diffuse sky regions, the main difference between emission 
predictions from these two models will be overall offsets due to differing 
input map zero levels.

\subsection{The 100-217 GHz Frequency Range}
\label{sec:lofreq}
FDS99 originally performed their FIRAS+DIRBE dust SED analysis for the sake of
accurately forecasting low-frequency CMB foregrounds. Recently, Galactic CMB
foregrounds, especially in the 100-150 GHz frequency range, have become a focal
point of cosmology owing to the \cite{bicep2} $B$-mode polarization results. 
Here we show that our two-component foreground predictions remain accurate on 
average to within $2.2$\% from 100-217 GHz, and we quantify the benefit of
using our two-component emission predictions in this frequency range relative
to extropolating the \cite{planckdust} single-MBB model.

\begin{figure*}
\begin{center}
\epsfig{file=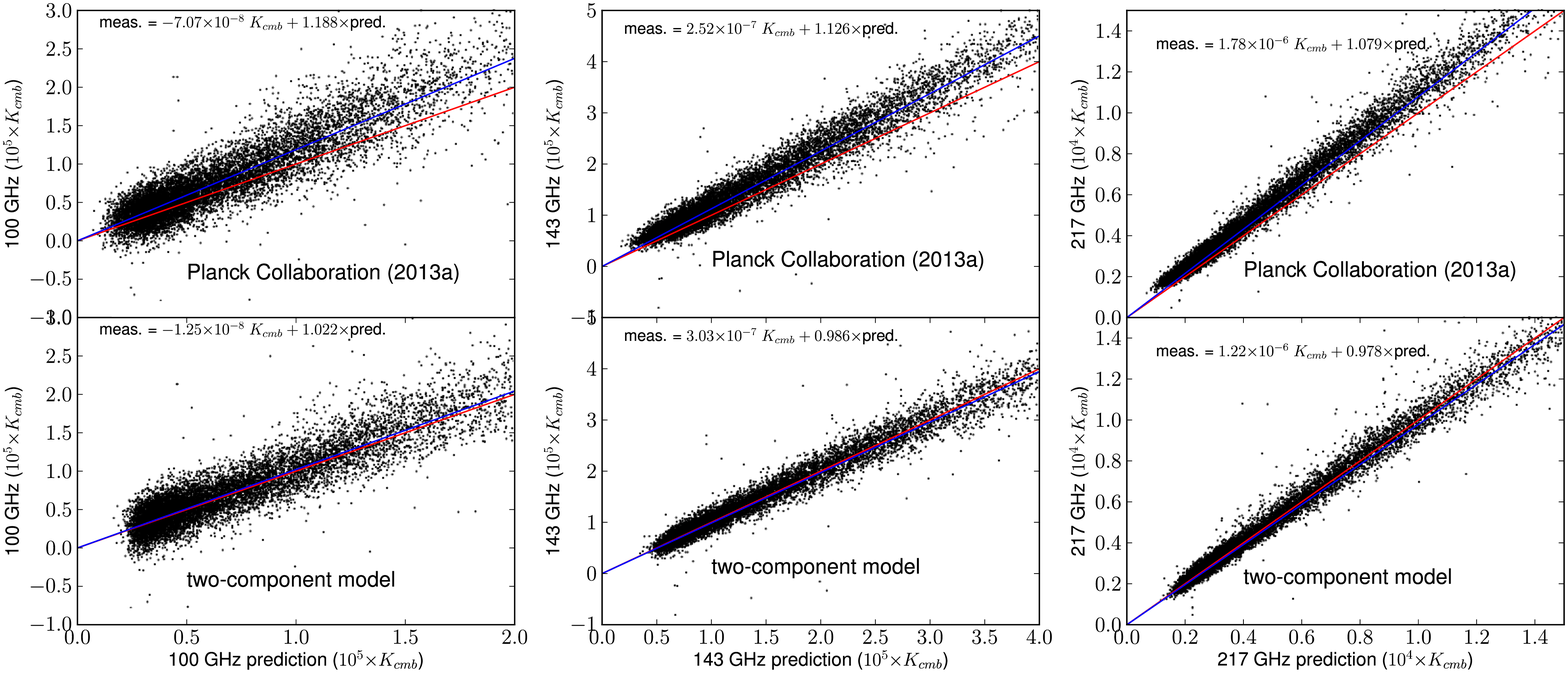, width=6.8in}
\caption{\label{fig:lofreq} Comparison between low-frequency thermal dust 
emission predictions from our best-fit two-component model (Table 
\ref{tab:global}, model 2) and those based on extrapolation of the 
\cite{planckdust} model. The top row shows scatter plots of the 
\cite{planckdust} predictions versus observed \PLANCK~100 GHz (left), 
\PLANCK~143 GHz (center) and \PLANCK~217 GHz (right). The bottom row shows 
scatter plots of the corresponding two-component predictions versus 
\PLANCK~observations. In all cases, the blue line indicates the best-fit linear
relationship, while the red line represents a perfect match between predictions
and observations. The lines of best-fit illustrate that the single-MBB model 
systematically underpredicts emission (in the multiplicative sense) by 18.8\%, 
12.6\% and 7.9\% at 100, 143 and 217 GHz respectively. On the other hand, by 
the same metric, the two-component model predictions at 100-217 GHz are always 
accurate to within $\le$2.2\%. The two-component fit results shown are based on
217-3000 GHz observations, meaning that the 100 GHz and 143 GHz predictions are
truly extrapolations, while the 217 GHz agreement is enforced by the fitting 
process itself to some extent.}
\end{center}
\end{figure*}

To assess the accuracy of low-frequency emission predictions, we compare the 
observed \PLANCK~HFI map at each of 100, 143, 217 GHz to the corresponding 
single-MBB and two-component predictions, with all maps smoothed to $1^{\circ}$
FWHM and binned down to $N_{side}$=$64$. We restrict to the same set of pixels 
used for the goodness-of-fit analysis of $\S$\ref{sec:hier}, with 
$|b|>30^{\circ}$ and $|\beta|>10^{\circ}$, also avoiding molecular emission, 
the SMICA inpainting mask, and compact sources. We then perform a linear fit 
between the \PLANCK~observed emission and the predicted emission at each 
frequency and for each emission model. For these fits, we consider the 
predicted emission to be the independent variable, since it has higher S/N than
the observations, especially at 100 and 143 GHz. We also assign pixel weights 
proportional to the predicted emission, so that the best-fit lines faithfully 
capture the linear trend exhibited without being biased by the large number of 
very low S/N pixels with minimal emission. Scatter plots between the predicted 
and observed emission are shown in Figure \ref{fig:lofreq}. The best-fit lines 
are overplotted and their equations are given in the top left corner of each 
subplot.

In both the single-MBB and two-component cases, all of the best fit offsets are
within the uncertainties quoted in Table \ref{table:offs}. On the other hand, 
the top row of Figure \ref{fig:lofreq} shows that the \cite{planckdust} 
single-MBB extrapolations yield slopes substantially different from unity: 
1.079 at 217 GHz, 1.126 at 143 GHz, and 1.188 at 100 GHz. The fact that the 
slopes are larger than unity indicates that the \cite{planckdust} 
extrapolations are systematically low. The systematic underprediction evidently
becomes gradually more pronounced as lower frequencies are considered, with a 
7.9\% underprediction at 217 GHz, a 12.6\% underprediction at 143 GHz and an 
18.8\% underprediction at 100 GHz. A deficit in single-MBB predictions relative
to the observed \PLANCK~100-217 GHz emission was also noted in 
\cite{planckdust2011}, e.g. their Figure 7.

For the case of the two-component model, we perform full-resolution 217-3000 
GHz fits using the \PLANCK+DIRBE favored global parameters (Table 
\ref{tab:global}, model 2), then smooth to $1^{\circ}$ FWHM and bin down to 
$N_{side}=64$ before predicting the 100-217 GHz emission. The bottom row of
Figure 14 shows that each of the best-fit lines is very similar to the 
corresponding red line which represents a perfect match between predicted and 
observed emission. More quantitatively, the two-component slopes are all within
2.2\% of unity: 0.978 at 217 GHz, 0.986 at 143 GHz and 1.022 at 100 GHz. We 
note that at 217 GHz, the good agreement is in some sense predetermined by the 
fact that \PLANCK~217 GHz has been included in our two-component MCMC fits. On 
the other hand, the 143 and 100 GHz predictions are based on extrapolation.

We conclude from these predicted versus observed emission comparisons that our 
two-component model outperforms extrapolation of the \cite{planckdust} 
single-MBB model at predicting Galactic thermal dust emission in diffuse 
regions from 100-217 GHz. It should be reiterated, once again, that 
\cite{planckdust} did not intend for their single-MBB model to be extrapolated 
to frequencies below 350 GHz (see their $\S$7.2.1), whereas we optimized our 
two-component model to be valid over the entire 100-3000 GHz frequency range.
Our two-component model thus represents the first \PLANCK~based thermal
dust emission model valid over the entire 100-3000 GHz frequency range.

\section{Data Release}
\label{sec:release}
\begin{figure*}
\begin{center}
\epsfig{file=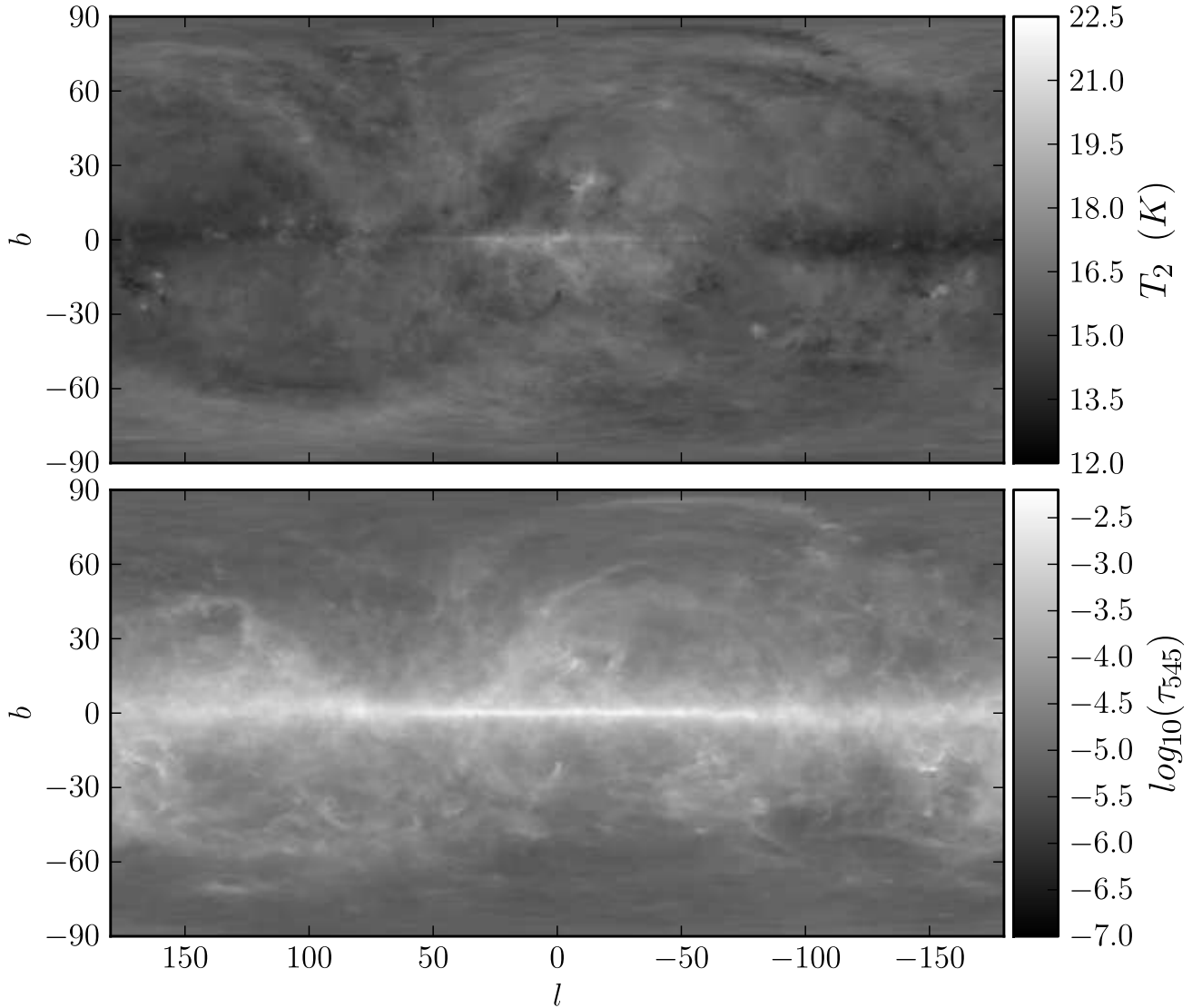, width=7.0in}
\caption{\label{fig:results} (top) Hot dust temperature derived from our 
full-resolution two-component model fits of \PLANCK~217-857 GHz and SFD 
100$\mu$m, downbinned to 27.5$'$ resolution. (bottom) Corresponding full-sky 
map of best-fit two-component 545 GHz optical depth.}
\end{center}
\end{figure*}

We are releasing a set of $N_{side}$=2048 HEALPix maps in Galactic coordinates 
which summarize the results of our full-resolution two-component dust fits. 
Low-resolution renderings of our full-sky dust temperature and optical depth 
maps are shown in Figure \ref{fig:results}. Our data release also includes
software utilities for obtaining emission and reddening predictions from our 
\PLANCK-based two-component fits. Refer to the data release documentation and 
FITS file headers for further details.\footnote{\texttt{http://faun.rc.fas.harvard.edu/ameisner/planckdust}}

\section{Conclusions}
\label{sec:conclusion}

\subsection{Single-MBB versus Two-component emission}
A major aim of this work has been to determine whether the  FDS99 two-component
dust emission model remains favored over single-MBB models when swapping the 
\PLANCK~HFI maps for FIRAS at frequencies below 1250 GHz. We compared dust SED 
models in two ways (1) by fitting a 100-3000 GHz spectrum composed of per-band 
correlation slopes versus \PLANCK~857 GHz (2) by finding the best-fit dust 
temperature and optical depth per line-of-sight, with each pixel's SED 
comprised of 100-3000 GHz \PLANCK+DIRBE data, and comparing the average 
goodness-of-fit under various emission models.

In both the correlation slope analysis of $\S$\ref{sec:global} and the 
goodness-of-fit analysis of $\S$\ref{sec:hier} we found that the best-fit 
\PLANCK+DIRBE two-component model (Table \ref{tab:global}, model 2) 
outperformed the best-fit single-MBB model, but by a lesser margin in 
$\chi^2_{\nu}$ than found by FDS99 using FIRAS+DIRBE. Specifically, our 
best-fit \PLANCK+DIRBE two-component model yielded an improvement of 
$\Delta \chi^2_{\nu}$=3.41 ($\S$\ref{sec:global}) and $\Delta \chi^2_{\nu}$=0.4
($\S$\ref{sec:hier}). This represents a far less dramatic contrast in 
$\chi^2_{\nu}$ than found by the FDS99 correlation slope analysis, 
$\Delta \chi^2_{\nu}$=29.2. Perhaps a relative lack of discrimination amongst 
competing dust SED models when relying on \PLANCK+DIRBE is to be expected, 
given that our constraints include only nine broad frequency channels, whereas 
FDS99 employed $>$200 narrow bands. Still, $\Delta \chi^2_{\nu}$=0.4 from 
$\S$\ref{sec:hier} is formally of enormous significance, given the $\sim$75,000
degrees of freedom in that analysis.

Nevertheless, we have established that the two-component emission model remains
viable in light of the \PLANCK~HFI data, and that the FIR/submm dust SED's 
preference for two MBB components rather than just one is not simply an 
idiosyncracy of the FIRAS spectra. Furthermore, we showed in 
$\S$\ref{sec:lofreq} that our 100-217 GHz two-component emission predictions 
are on average accurate to within 2.2\%, whereas extrapolating the 
\cite{planckdust} single-MBB model systematically underestimates low-frequency 
dust emission by 18.8\% at 100 GHz, 12.6\% at 143 GHz and 7.9\% at 217 GHz. We 
therefore recommend that those interested in thermal dust foregrounds in the 
100-3000 GHz frequency range use our data release to predict unpolarized dust 
emission, at the very least in order to help determine the level at which the 
choice of dust emission model may influence their conclusions.

\subsection{Towards a Replacement for SFD}
\label{sec:replace}
Because of the broad frequency coverage and high angular resolution afforded
by the \PLANCK~HFI full-sky maps, we initially speculated that a \PLANCK~based 
extinction map might easily outperform SFD, the most commonly used optical
reddening map. However, at this point in time, we do not yet recommend that the
results presented in this work be considered a replacement for SFD in terms of 
optical extinction/reddening estimates.

The CIBA remains a major imperfection that still requires further 
investigation. The CIB anisotropies are very evident in low-dust regions of our
maps of optical depth and predicted dust emission. As described in 
$\S$\ref{sec:mcmc}, we have propagated the CIBA RMS amplitudes and 
inter-frequency covariances into our uncertainty estimates through the 
likelihood function in our MCMC procedure. However, this treatment falls far 
short of actually removing the spatial imprint of the CIBA on our derived 
parameters. The CIB anisotropies are more prominent in our optical depth map 
relative to that of SFD because of the lower-frequency \PLANCK~maps we rely 
upon to achieve a high-resolution temperature correction.

Imperfect zodiacal light (zodi) corrections represent a second major limitation
of our results. The ecliptic plane's prominence in our full-sky temperature map
(Figure \ref{fig:results}) suggests that the zodiacal light subtractions 
performed on the input maps are not ideal. Our comparisons of the FIR maps used
in this study against H\,\textsc{i} emission bear out this notion, further 
revealing that the imperfect zodi corrections are not limited to \verb|i100|, 
but in fact are noticeable in all of the HFI \verb|R1.10_nominal_ZodiCorrected|
maps as well. We deemed it infeasible to reconsider all of the \PLANCK~zodi 
corrections in addition to the 3000 GHz zodi correction as a part of this 
study, especially considering that the forthcoming \PLANCK~2014 release is 
expected to include a revised/improved zodi subtraction.

Irrespective of the notable imperfections in our results, more detailed
comparisons between our reddening estimates here and those of SFD are required 
to determine/quantify which map is superior in particular applications. One 
definitive improvement of our reddening estimates relative to those of SFD is 
our ability to quote reddening uncertainties, which results from the 
probabilistic framework of $\S$\ref{sec:mcmc}. The extinction estimates from 
this work can also be employed as an alternative to those of SFD, to gauge the 
impact of dust map choice in a specific end user's application.

We thank the anonymous referee for helpful suggestions. We gratefully 
acknowledge support from the National Science Foundation Graduate Research 
Fellowship under Grant No. DGE1144152, and NASA grant NNX12AE08G. Based on 
observations obtained with Planck (http://www.esa.int/Planck), an ESA science 
mission with instruments and contributions directly funded by ESA Member 
States, NASA, and Canada. This research made use of the NASA Astrophysics Data 
System (ADS) and the IDL Astronomy User's Library at Goddard. \footnote{Available at \texttt{http://idlastro.gsfc.nasa.gov}}

\bibliographystyle{apj}
\bibliography{twocomp.bib}

\end{document}